\documentclass[twocolumn,journal]{IEEEtran}

\usepackage{color}
\usepackage{graphicx}
\usepackage{amsmath}
\usepackage{array}
\usepackage{url}
\usepackage{amssymb}
\usepackage{algorithm}
\usepackage{algorithmic}
\usepackage{amsmath}
\usepackage{multirow}
\usepackage{booktabs}
\usepackage{array}
\usepackage{amsthm}
\usepackage{caption}
\usepackage{stfloats}
\usepackage{subfigure}
\usepackage{bm}
\usepackage{booktabs}
\usepackage{setspace}
\usepackage{upgreek}
\usepackage{setspace}
\usepackage{diagbox}
\usepackage{gensymb}

\newcommand{\be}{\begin{equation}}
\newcommand{\ee}{\end{equation}}
\newcommand{\bea}{\begin{eqnarray}}
\newcommand{\eea}{\end{eqnarray}}
\newcommand{\ba}{\begin{array}}
\newcommand{\ea}{\end{array}}

\newcommand{\non}{\nonumber}

\allowdisplaybreaks[4]

\title{Dual-Functional Radar-Communication Waveform Design: A Symbol-Level Precoding Approach
\thanks{R. Liu and M. Li are with the School of Information and Communication Engineering, Dalian University of Technology, Dalian 116024, China (e-mail: liurang@mail.dlut.edu.cn; mli@dlut.edu.cn).}
\thanks{Q. Liu is with the School of Computer Science and Technology, Dalian University of Technology, Dalian 116024, China (e-mail: qianliu@dlut.edu.cn).}
\thanks{A. L. Swindlehurst is with the center for Pervasive Communications and Computing, University of California, Irvine, CA 92697, USA (e-mail: swindle@uci.edu).}
}

\author{Rang Liu,~\IEEEmembership{Graduate Student Member,~IEEE,}
        Ming Li,~\IEEEmembership{Senior Member,~IEEE,}
        Qian Liu,~\IEEEmembership{Member,~IEEE,}\\
        and A. Lee Swindlehurst,~\IEEEmembership{Fellow,~IEEE}}

\begin{document}

\maketitle

\pagestyle{empty}
\thispagestyle{empty}

\begin{abstract}
Dual-functional radar-communication (DFRC) systems can simultaneously perform both radar and communication functionalities using the same hardware platform and spectrum resource.
In this paper, we consider multi-input multi-output (MIMO) DFRC systems and focus on transmit beamforming designs to provide both radar sensing and multi-user communications.
Unlike conventional block-level precoding techniques, we propose to use the recently emerged symbol-level precoding approach in DFRC systems, which provides additional degrees of freedom (DoFs) that guarantee preferable instantaneous transmit beampatterns for radar sensing and achieve better communication performance.
In particular, the squared error between the designed and desired beampatterns is minimized subject to the quality-of-service (QoS) requirements of the communication users and the constant-modulus power constraint.
Two efficient algorithms are developed to solve this non-convex problem on both the Euclidean and Riemannian spaces.
The first algorithm employs penalty dual decomposition (PDD), majorization-minimization (MM), and block coordinate descent (BCD) methods to convert the original optimization problem into two solvable sub-problems, and iteratively solves them using efficient algorithms.
The second algorithm provides a much faster solution at the price of a slight performance loss, first transforming the original problem into Riemannian space, and then utilizing the augmented Lagrangian method (ALM) to obtain an unconstrained problem that is subsequently solved via a Riemannian Broyden-Fletcher-Goldfarb-Shanno (RBFGS) algorithm.
Extensive simulations verify the distinct advantages of the proposed symbol-level precoding designs in both radar sensing and multi-user communications.
\end{abstract}


\maketitle

\section{Introduction}
\vspace{0.2 cm}

Spectrum sharing has been regarded as a promising solution for tackling the spectrum congestion problem in rapidly expanding wireless communication networks.
One of the most popular examples is spectrum sharing between radar and wireless communication systems, which has been widely investigated from theoretical performance analyses to practical waveform designs \cite{Zheng SPM 2019}-\cite{Liu TCOM 2020}.
Previous work on joint radar-communication design mainly focuses on: 1) Radar-Communication Coexistence (RCC) and 2) Dual-Functional Radar-Communication (DFRC) designs.
In RCC systems, non-colocated radar and communication systems must exchange necessary side-information for performing interference management and achieving better cooperation \cite{Zheng JSTSP 2018}, \cite{Liu TSP 18}, which greatly increases the system complexity and cost.
On the other hand, DFRC systems simultaneously perform both radar and communication functionalities using the same signals transmitted from a fully-shared transmitter, which only requires one smaller-size, lower-cost, and lower-complexity platform and naturally achieves full cooperation.
Therefore, DFRC systems have a competitive advantage and many novel applications have been proposed in recent years \cite{Zhang VTM 2021}.

In DFRC systems, the radar and communication functionalities inherently have conflicting requirements in terms of, e.g., the antenna placement, the operation region of the power-amplifiers, the signal formats, etc.
Therefore, the transmit waveform should be carefully designed to balance the requirements of these two functionalities and achieve better system performance. In addition, multi-input multi-output (MIMO) architectures are also widely applied in DFRC systems to provide waveform diversity for radar target detection \cite{Li SPM 2007}, and beamforming gains and spatial multiplexing for multi-user communications.

The existing research on waveform design for DFRC systems can be divided into two main categories: radar-centric and communication-centric approaches.
The radar-centric approach prioritizes radar sensing functionality and realizes communication by embedding information symbols into the radar waveform \cite{Wu TCOM 2020,Hassanien TSP 2016}, which permits only a very low transmission rate since only limited symbols can be embedded into each radar pulse.
Furthermore, considering security and cost, government and military agencies usually do not allow changes to their radar systems \cite{Zhang VTM 2021}.
Thus the communication-centric approach, which typically relies on transmit beamforming from multi-antenna base stations (BSs) to support radar sensing, is more attractive and more widely considered.

Recently, many researchers have devoted themselves to transmit beamforming designs in MIMO DFRC systems \cite{McCormick IRC 2017}-\cite{Su TWC 2020}, where the precoding matrix is optimized with different radar sensing and communication metrics.
Typical radar sensing metrics include the radar receiver's signal-to-interference-plus-noise ratio (SINR) \cite{Qian TSP 2018}, the beampattern mean squared error (MSE) \cite{Tang SAM 2020}, the Cram\'{e}r-Rao bound \cite{Kumari TSP 2020}, and the similarity between the designed beamformer and that of the reference radar-only system \cite{Liu WCL 2020}-\cite{Cheng SAM 2020}.
Meanwhile, widely-used  communication metrics include the achievable rate \cite{Yuan TWC 2020}, \cite{Xu Access 2020}, the communication user's SINR \cite{Liu WCL 2020}, \cite{Liu TWC 18}, \cite{Su TWC 2020}, and the multi-user interference (MUI) \cite{Tang SAM 2020}, \cite{Liu TSP 2018}.
The combination of radar sensing and communication metrics can provide a comprehensive criteria for evaluating the performance of DFRC systems \cite{Tang SAM 2020}-\cite{Liu TSP 2018}.

In the transmit waveform designs mentioned above \cite{McCormick IRC 2017}-\cite{Su TWC 2020}, conventional linear block-level precoding is used to embed the communication symbols into the dual-function transmit waveform. The available degrees of freedom (DoFs) of these approach have been proven to be limited by the number of users \cite{Liu TSP 20}.
In order to increase the number of DoFs in the transmit waveform for better radar sensing performance, the authors in \cite{Liu TSP 20}-\cite{Liu RadarConf 2020} proposed to transmit both the precoded communication symbols and the radar waveform, which are jointly designed in a block-level fashion.
However, such block-level precoding designs inherently have limited DoFs due to their restriction to linear processing.
More importantly, since block-level precoding designs employ performance metrics based on second-order statistics (e.g., SINR and MSE) to optimize the average transmit beampattern, the radar sensing performance can be guaranteed only when the number of transmitted symbols is sufficiently large.
In other words, the instantaneous transmit beampatterns in different time slots might have significant distortions, which causes severe performance degradation on target detection and parameter estimation if only a limited number of samples are collected.
In light of these shortcomings for block-level precoding, more sophisticated transmit beamforming strategies are necessary to fully exploit the DoFs available for simultaneously providing reliable radar sensing and high-rate communications.
This finding motivates us to employ the recently emerged symbol-level precoding approach in DFRC systems.

Symbol-level precoding has been proposed as a way of exploiting rather than simply eliminating MUI in multi-user communication systems \cite{Masouros TWC 2009}-\cite{Liu TVT 2021}.
Unlike conventional block-level precoding, symbol-level precoding is a non-linear and symbol-dependent approach, which optimizes each instantaneous transmitted vector based on the specific symbols to be transmitted. In this way, the instantaneous transmit beampattern in each time slot can be carefully designed in a symbol-by-symbol fashion to provide more DoFs for radar sensing functionality.
From the communication perspective, instead of suppressing MUI in a statistical manner, symbol-level precoding can exploit the transmitted symbol information to convert MUI into constructive components for improving the quality-of-service (QoS) of multi-user communications.

In prior work \cite{Liu TSP 18}, symbol-level precoding technique has been employed in a RCC system to enjoy the advantages of interference exploitation.
However, in RCC systems, the symbol-level precoding design only optimizes the communication functionality and simultaneously suppresses the interference to the radar system regardless of specific radar waveforms.
On the contrary, in the considered DFRC system in this paper, the symbol-level precoding design focuses on the dual-use waveform, which simultaneously realizes both communication and radar sensing functionalities.
Therefore, the symbol-level precoding design for DFRC systems is quite different from that for RCC systems, and has not been previously investigated in the literature.

In this paper, we investigate symbol-level precoding designs for a MIMO DFRC system, where a multi-antenna BS simultaneously serves multiple single-antenna communication users and detects targets from several directions of interest. In particular, the symbol-level precoded transmit vector is optimized to minimize the squared error between the obtained and desired beampatterns under the communication QoS requirements and the constant-modulus power constraint.
The main contributions can be summarized as follows:

\begin{itemize}
  \item For the first time, we employ symbol-level precoding techniques in DFRC systems to provide additional DoFs for the waveform designs, which (i) allows desirable transmit beampatterns to be realized in every time slot rather than just ``on average'', and (ii) significantly improves the simultaneous multi-user communication performance by converting harmful MUI into useful signals.
      Compared with conventional designs based on second-order statistics and block-level precoding, our symbol-level precoding approach imposes different radar sensing and multi-user communication constraints and leads to a brand-new waveform design problem.

  \item In order to handle the resulting complicated non-convex waveform design problem, we first develop an algorithm framework that combines penalty dual decomposition (PDD), majorization-minimization (MM), and block coordinate descent (BCD) methods to convert the problem into two solvable sub-problems. Then, a closed-form phase alignment and a Lagrangian dual approach are applied to efficiently solve these sub-problems.

  \item We further propose a more computationally efficient solution that results in only a slight performance loss. In this second approach, we convert the original problem into Riemannian space, and utilize the augmented Lagrangian method (ALM) to transform it into an unconstrained problem. Then, an efficient Riemannian Broyden-Fletcher-Goldfarb-Shanno (RBFGS) algorithm is developed to solve this problem.

  \item Finally, we provide extensive simulation results to demonstrate the distinct advantages of the proposed symbol-level precoding designs in both radar sensing and multi-user communications.
\end{itemize}

The rest of this paper is organized as follows.
Section \ref{sec:system model} introduces the system model, the performance metrics for multi-user communications and radar sensing, and the problem formulation.
The proposed PDD-MM-BCD and ALM-RBFGS algorithms are developed in Sections \ref{sec:Algorithm 1} and \ref{sec:Algorithm 2}, respectively.  Possible extensions of the proposed approach to scenarios that require Doppler processing are described in Section~V. Simulation results are presented in Section \ref{sec:simulation results}, and finally conclusions are provided in Section \ref{sec:conclusions}.

\textit{Notation}:
Boldface lower-case and upper-case letters indicate column vectors and matrices, respectively.
$(\cdot)^T$ and $(\cdot)^H$ denote the transpose and the transpose-conjugate operations, respectively.
$\mathbb{C}$ denotes the set of complex numbers.
$| a |$ and $\| \mathbf{a} \|$ are the magnitude of a scalar $a$ and the norm of a vector $\mathbf{a}$, respectively.
$\angle{a}$ is the angle of complex-valued $a$.
$\mathfrak{R}\{\cdot\}$ and $\mathfrak{I}\{\cdot\}$ denote the real and imaginary part of a complex number, respectively.
$\odot$ denotes the Hadamard product.
$\mathbf{A} \succeq \mathbf{0}$ indicates that the matrix $\mathbf{A}$ is positive semi-definite.
$\mathbf{I}_M$ indicates an $M\times M$ identity matrix.
Finally, we adopt the following indexing notation: $\mathbf{A}(i,j)$ denotes the element of the $i$-th row and the $j$-th column of matrix $\mathbf{A}$, and $\mathbf{a}(i)$ denotes the $i$-th element of vector $\mathbf{a}$.

\section{System Model and Problem Formulation}\label{sec:system model}
\vspace{0.2 cm}

\begin{figure}[!t]
\centering
\includegraphics[width = 2.2 in]{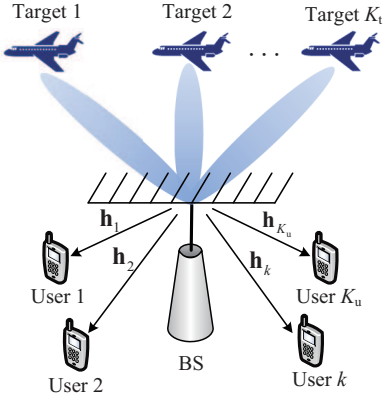}\vspace{0.1 cm}
\caption{A dual-functional radar-communication system.}
\label{fig:system_model}\vspace{0.2 cm}
\end{figure}

\subsection{System Model}
%\vspace{0.1 cm}

We consider a colocated monostatic MIMO DFRC system as shown in Fig. 1, where the BS is equipped with $M$ antennas in a uniform linear array (ULA).
The BS simultaneously serves $K_\text{u}$ single-antenna communication users and detects the locations of $K_\text{t}$ targets.
Generally, $K_\text{u} \leq M$ and $K_\text{t} \leq M$.
The BS periodically emits short and high-power pulse-modulated signals to implement DFRC.
The same antenna array is used for both transmit and receive in different time slots via time-division (TD) processing.
In particular, the BS first transmits radar pulses that are embedded with communication symbols to simultaneously illuminate the targets of interest and transfer information to the communication users.
Then, the BS switches to the radar receiver mode, collects the reflected echo signals from the targets, and further estimates the targets' parameters of interest based on the prior knowledge of the transmitted radar pulses and advanced estimation algorithms.

In this paper, we employ non-linear symbol-level precoding to realize both multi-target radar sensing and multi-user communications.
Specially, let $\mathbf{x}[n] \triangleq \left[x_1[n], x_2[n], \ldots, x_M[n]\right]^T$ be the precoded transmit vector in the $n$-th time slot (symbol duration), where $x_m[n]$ is the baseband signal transmitted from the $m$-th antenna, $m = 1, 2, \ldots, M$.
Let $\mathbf{s}[n]\triangleq \left[s_1[n], s_2[n], \ldots, s_{K_\text{u}}[n]\right]^T$ be the symbols  transmitted to the $K_\text{u}$ users.
Unlike conventional block-level precoding designs where $\mathbf{x}[n]$ is a linear function of $\mathbf{s}[n]$, symbol-level precoding in general employs a non-linear mapping from $\mathbf{s}[n]$ to $\mathbf{x}[n]$, and optimizes $\mathbf{x}[n]$ directly according to the instantaneous symbol vector $\mathbf{s}[n]$ instead of using second-order statistics-based metrics.
Therefore, symbol-level precoding can exploit more DoFs to improve both radar sensing and multi-user communication performance, as well as guarantee superior instantaneous transmit beampatterns.
Since the conventional statistical metrics, e.g., SINR and MSE, are inappropriate to evaluate the performance of the symbol-level precoding designs, we will describe the symbol-level precoding performance metrics for multi-user communications and radar sensing in the following two subsections, respectively.

\subsection{Multi-user Communication Performance Metric}
%\vspace{0.1 cm}

 To simplify the description of our approach, in this paper we will
assume that the transmitted symbols $\mathbf{s}[n]$ are independently selected from an $\Omega$-phase-shift-keying (PSK) constellation.  Modifications necessary to accommodate other modulation types, e.g., quadrature amplitude modulation (QAM), will be briefly discussed below.
For transmitting $\mathbf{s}[n]$ in the $n$-th time slot, the symbol-level precoded transmit vector  $\mathbf{x}[n]$ is designed and transmitted from the multiple antennas.
The received signal at the $k$-th user is thus written as
\begin{equation}
r_k[n] = \mathbf{h}_k^H\mathbf{x}[n] + n_k[n],
\end{equation}
where $\mathbf{h}_k\in \mathbb{C}^M$ is the Rayleigh fading channel from the BS to the $k$-th user, and $n_k[n]\sim \mathcal{CN}\left(0,\sigma_k^2\right)$ is the additive white Gaussian noise (AWGN) at the $k$-th user.

\begin{figure}[!t]
\centering
\subfigure[An example of the constructive region.]{
\begin{minipage}{4 cm}
\centering
\includegraphics[width = 0.85\textwidth]{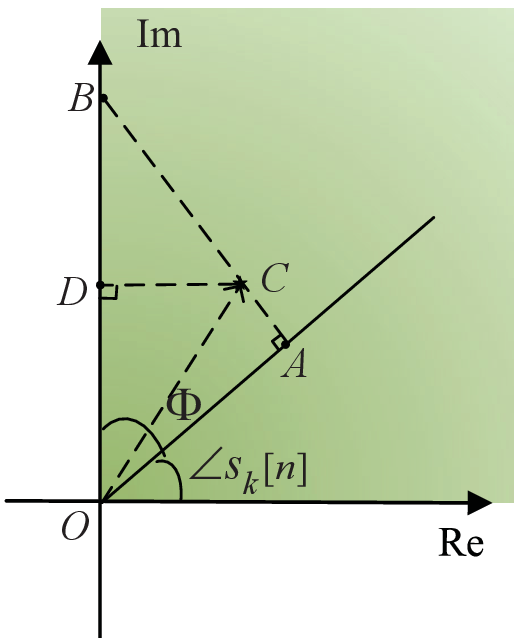}
\vspace{0.8 cm}
\label{fig:CR1}
\end{minipage}
}
\subfigure[After rotating the diagram in Fig. \ref{fig:CR1} clockwise.]{
\begin{minipage}{4 cm}
\centering
\includegraphics[width = 0.7\textwidth]{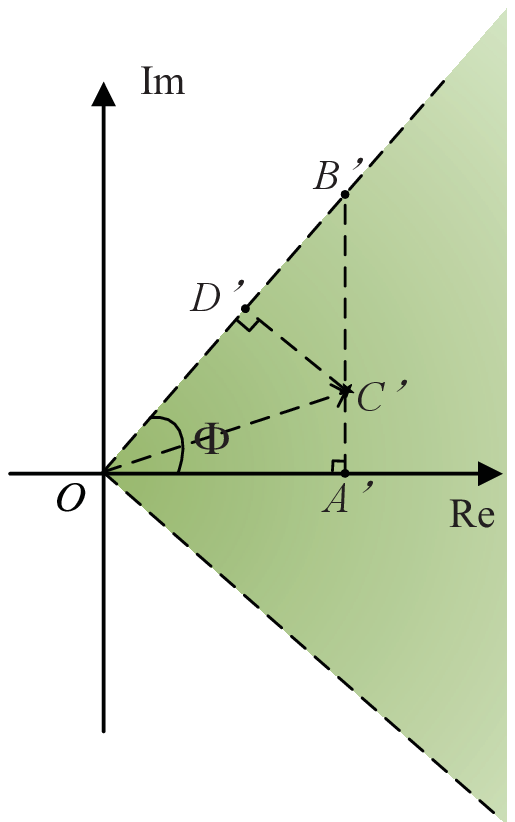}
\vspace{0.5 cm}
\label{fig:CR2}
\end{minipage}
}
\caption{Constructive region for a QPSK symbol.}
\label{fig:CR}\vspace{0.2 cm}
\end{figure}

Symbol-level precoding relies on the idea of constructive interference (CI), in which harmful MUI is converted into helpful signals that push the received signals farther away from their decision boundaries \cite{Masouros TWC 2009}-\cite{Liu TVT 2021}.
Thus, the Euclidean distance between the received noise-free signal and its closest decision boundaries is usually adopted as the QoS metric to evaluate the communication performance since it directly determines the symbol error rate (SER). The idea is illustrated in Fig. \ref{fig:CR} for
quadrature-PSK (QPSK) signals, where the green sector is the decision region when the desired symbol is $s_k[n] = e^{j\pi/4}$, $\overrightarrow{OC} = \mathbf{h}_k^H\mathbf{x}[n]$ is the received noise-free signal at the $k$-th user, $\Phi = \pi/\Omega$, points $A$ and $D$ are the projections of point $C$ on the direction of $s_k[n]$ and the corresponding  nearest decision boundary, respectively, and point $B$ is the intersection of the extension of $\overrightarrow{AC}$ and the nearest decision boundary.
We see that $s_k[n]$ can be correctly detected at the $k$-th user when the received signal $r_k[n]$ lies in the green region.
In order to improve the robustness to noise, the transmitted signal $\mathbf{x}[n]$ should be designed such that the received noise-free signal $\overrightarrow{OC}$ is as far away from its decision boundaries as possible.
Therefore, the minimum Euclidean distance between the received noise-free signal and its decision boundaries, i.e., $|\overrightarrow{CD}|$, is taken to be the communication QoS metric.
In order to facilitate the expression of $|\overrightarrow{CD}|$, we rotate the diagram in Fig. \ref{fig:CR1} clockwise by $\angle s_k[n]$ degrees as shown in Fig. \ref{fig:CR2}.
Then, the distance between the received noise-free signal and its closest decision boundary can be readily expressed as
\begin{equation}
\begin{array}{l}
\big|\overrightarrow{C'D'}\big| = \big|\overrightarrow{C'B'}\big|\cos\Phi = \big(\big|\overrightarrow{A'B'}\big|-\big|\overrightarrow{A'C'}\big|\big)\cos\Phi \\[6pt]
 = \big[\mathfrak{R}\{\overrightarrow{OC'}\}\tan\Phi -
|\mathfrak{I}\{\overrightarrow{OC'}\}|\big]\cos\Phi \\[6pt]
\label{eq:distance}
 = \mathfrak{R}\{\mathbf{h}_k^H\mathbf{x}[n]e^{-j\angle s_k[n]}\}\sin\Phi
\hspace{-0.1 cm} - \hspace{-0.1 cm} \left|\mathfrak{I}
\{\mathbf{h}_k^H\mathbf{x}[n]e^{-j\angle s_k[n]}\}\right|\cos\Phi.
\end{array}
\end{equation}

Thus, to guarantee the multi-user communication QoS, the transmitted signal $\mathbf{x}[n]$ should be designed to ensure that the distance (\ref{eq:distance}) is no less than a preset minimum; i.e., the communication constraint is formulated as
\begin{equation}\begin{aligned}\label{eq:fig2 communication constraint}
&\mathfrak{R}\big\{\mathbf{h}_k^H\mathbf{x}[n]e^{-j\angle{s_k[n]}}\big\}
   \sin \Phi \\
&~~~~~~~~~~~~~~~~~~- \big|\mathfrak{I}\big\{\mathbf{h}_k^H\mathbf{x}[n]e^{-j\angle{s_k[n]}}\big\}\big|\cos\Phi
   \geq \beta_k,
\end{aligned}\end{equation}
where $\beta_k > 0$ is the preset minimum QoS requirement for the $k$-th user.

 The approach described above is similar for other types of signal constellations; the constructive region (CR) for different constellations is just expressed by a different set of inequalities. For example, in the case of QAM, there are three different types of CRs, depending on whether the symbol is from an inner constellation point, an outer point on the corners, or an outer point on the sides. Each type of CR is a convex region that can be expressed as linear inequalities on $\mathfrak{R}\{\mathbf{h}_k^H\mathbf{x}[n]\}$ and $\mathfrak{I}\{\mathbf{h}_k^H\mathbf{x}[n]\}$. (See \cite{Liu ICC 2020}-\cite{JeddaMSN18} for further details). Thus, the algorithms proposed later in this work can be easily extended to QAM and indeed arbitrary modulation formats, provided the symbol decision regions are convex.

\subsection{Radar Sensing Performance Metric}
%\vspace{0.1 cm}

For radar sensing purposes, the waveforms are usually designed to direct the transmit beam towards the directions of potential targets, so that they can be illuminated by stronger signals.
This allows the radar receiver to obtain stronger echo signals from the targets, which yields more accurate estimation of the parameters of interest.
To evaluate the transmit beampattern, in this subsection we first derive the received signal and corresponding beampattern for each direction, and then formulate the squared error between the designed and desired beampatterns as a radar sensing performance metric.

In MIMO radar systems, narrow-band waveforms and line-of-sight (LoS) propagation are usually assumed.
Thus, the baseband signal at the angular direction $\theta\in\left[-\frac{\pi}{2}, \frac{\pi}{2}\right)$ can be expressed as
\begin{equation}
r\left(n;\theta\right) = \mathbf{a}^H(\theta)\mathbf{x}[n],
\end{equation}
where $\mathbf{a}(\theta)\triangleq\left[1,e^{j\frac{2\pi}{\lambda}\Delta\sin(\theta)}, \ldots, e^{j\frac{2\pi}{\lambda}(M-1)\Delta\sin(\theta)}\right]^T\in \mathbb{C}^M$ is the transmit steering vector for direction $\theta$, with $\Delta$ representing the antenna spacing and $\lambda$ the wavelength.
When the transmit signal is reflected by $K_\text{t}$ point-like targets at the directions $\theta_{k_\text{t}},~~k_\text{t} = 1,\ldots, K_\text{t}$, the received signal at the BS is written as
\begin{equation}
\mathbf{y}[n] = \sum_{k_\text{t}=1}^{K_\text{t}}\beta_{k_\text{t}}\mathbf{a}(\theta_{k_\text{t}})\mathbf{a}^H(\theta_{k_\text{t}})\mathbf{x}\left[n\right] + \mathbf{z}[n],
\end{equation}
where $\beta_{k_\text{t}}$ is the complex amplitude proportional to the radar-cross section (RCS) of the target at the direction $\theta_{k_\text{t}}$, and $\mathbf{z}[n] \sim \mathcal{CN}(0,\sigma^2_\text{z})$ is AWGN.
In this work, we assume that the targets are stationary or very slowly-moving with Doppler frequencies near zero, and they are located in the same range bin for simplicity.
Therefore, the radar sensing problem in this paper essentially focuses on estimating the RCS $\beta_{k_\text{t}}$ and the angular direction $\theta_{k_\text{t}}$ for all $k_\text{t}$.
This further implies that our problem formulation will not consider the temporal correlation properties of the transmitted waveforms. In Section~V we briefly discuss some ideas for extending the approach to the case with non-negligible Doppler.

To facilitate the target detection and parameter estimation, a widely acknowledged method is to maximize the signal power in the directions of potential targets and minimize it elsewhere, in such a way to enhance the echo signals from the targets and suppress clutter. In such approaches, the transmit beampattern is optimized to approach an ideal radar beampattern.
Thus, the similarity between the designed and desired transmit beampattern is a popular radar waveform design metric \cite{Li SPM 2007}, \cite{Stoica TSP 2007}.
It is noted that in existing works \cite{McCormick IRC 2017}-\cite{Liu RadarConf 2020} using conventional block-level precoding, the \textit{average} transmit beampattern is considered in the waveform optimization:
\begin{equation}\label{eq:radar-only beampattern}
P\left(\theta;\mathbf{R}\right) = \mathbb{E}\left\{\left|\mathbf{a}^H(\theta)\mathbf{x}[n]\right|^2\right\} = \mathbf{a}^H(\theta)\mathbf{R}\mathbf{a}(\theta),
\end{equation}
where $\mathbf{R} \triangleq \mathbb{E}\left\{\mathbf{x}[n]\mathbf{x}[n]^H\right\}$ is the covariance matrix of the transmitted signals.
In a block-level precoding system with precoding matrix $\mathbf{F}$ and $\mathbf{x}[n] = \mathbf{F}\mathbf{s}[n]$, the transmitted symbols are typically assumed to be statistically independent, i.e., $\mathbb{E}\{\mathbf{s}[n]\mathbf{s}[n]^H\} = \mathbf{I}_{K_\text{u}}$.
Thus, the design of the covariance matrix of the transmitted signals is equal to that of the precoder, i.e., $\mathbf{R} = \mathbf{F}\mathbf{F}^H$.
In designing the precoder $\mathbf{F}$ to match the ideal beampattern, since the second-order statistics of the transmitted symbols are used to derive the average transmit beampattern (\ref{eq:radar-only beampattern}), the number of transmitted signals/waveform samples must be sufficiently large to support this assumption.
However, in DFRC systems, the number of waveform samples in each radar pulse is usually limited, which consequently might cause significant distortions to the actual average beampattern obtained with only a few radar pulses. This can result in severe target detection and parmeter estimation perfomance since the transmit power cannot be better focused on the directions of interest to obtain stronger reflected echoes from the targets for sensing.
In order to overcome this drawback, we employ symbol-level precoding, which optimizes the {\em instantaneous} transmit beampattern by directly designing the transmitted signal $\mathbf{x}[n]$, and thus can provide satisfactory radar sensing performance with a limited number of samples.

With the transmitted signal $\mathbf{x}[n]$ in the $n$-th time slot, the instantaneous transmit beampattern (signal power) at direction $\theta$ is given by
\begin{equation}\label{eq: instantaneous beampattern}
P\left(\theta;\mathbf{x}[n]\right) = \left|\mathbf{a}^H(\theta)\mathbf{x}[n]\right|^2 = \mathbf{x}^H[n]\mathbf{A}(\theta)\mathbf{x}[n],
\end{equation}
where $\mathbf{A}(\theta) \triangleq \mathbf{a}(\theta)\mathbf{a}^H(\theta)$ for brevity.
Then, the radar sensing performance metric, i.e., the difference between the designed instantaneous transmit beampattern $P(\theta;\mathbf{x}[n])$ and desired beampattern $d(\theta)$, is formulated in terms of squared error as
\begin{equation}\label{eq: objective}
f\left(\alpha,\mathbf{x}[n]\right) = \frac{1}{L}\sum_{l=1}^L\left|\alpha d(\theta_l)-\mathbf{x}^H[n]\mathbf{A}(\theta_l)\mathbf{x}[n]\right|^2,
\end{equation}
where $\alpha$ is a scaling factor and $\left\{\theta_l\right\}_{l=1}^L$ are the sampled angles.

\subsection{Problem Formulation}
%\vspace{0.1 cm}

Based on the above problem description, in this paper we aim to design the symbol-level precoded transmit vector $\mathbf{x}[n]$ to minimize the difference between the designed and desired transmit beampatterns, while satisfying the CI-based QoS requirements of the multi-user communication system and the transmit power constraint.

It is noted that, in practical implementations, each transmit antenna uses its maximal available power to transmit the radar waveform in order to achieve the highest power efficiency \cite{Stoica TSP 2007}, \cite{Sun ICASSP 2015}.
Thus, each element of the transmitted signal $\mathbf{x}[n]$ has a constant-modulus power constraint, i.e.,
\begin{equation}\label{eq: power constraint}
\left|x_m[n]\right| = \sqrt{P_\text{tot}/{M}},~~\forall m = 1, 2, \ldots, M,
\end{equation}
where $P_\text{tot}$ is the maximum total transmit power.
It should be emphasized that, for the symbol-level precoding approach, this constant-modulus power constraint holds at the symbol level (i.e., for each time slot), and thus guarantees an extraordinarily low peak-to-average power ratio (PARR), which is earnestly pursued in commercial radar systems employing low-cost and non-linear amplifiers.

For conciseness, we drop the time slot index $[n]$ in the the rest of the paper.
Therefore, the optimization problem is formulated as
\begin{subequations}
\label{eq:original problem}
\begin{align}
\label{eq:original objective}
&\underset{\mathbf{x},\alpha}{\min}~~\frac{1}{L}\sum_{l=1}^L\left|\alpha d(\theta_l)-\mathbf{x}^H\mathbf{A}(\theta_l)\mathbf{x}\right|^2 \\
\label{eq:communication constraint}
&~\text{s.t.}~~\mathfrak{R}\left\{\mathbf{h}_k^H\mathbf{x}e^{-j\angle{s_k}}\right\}
   \sin \Phi \non \\
   &~~~~~~~~~~~~~~~~~ - \left|\mathfrak{I}\left\{\mathbf{h}_k^H\mathbf{x}e^{-j\angle{s_k}}\right\}\right|\cos\Phi
   \geq \beta_k,~~\forall k,\\
\label{eq:unit modulus constraint}
&~~~~~~~\left|x_m\right| = \sqrt{P_\text{tot}/M},~~\forall m.
\end{align}
\end{subequations}
Before the algorithm development, we first re-formulate this problem in a more compact format.
It is obvious that the original problem (\ref{eq:original problem}) is a quadratic function in the variable $\alpha$.
Thus, the minimum of $f(\alpha,\mathbf{x})$ is achieved when
\begin{equation}
\hspace{-0.07 cm}\frac{\partial f(\alpha,\mathbf{x})}{\partial\alpha} = \frac{1}{L}\sum_{l=1}^L\left[2\alpha d^2(\theta_l)-2d(\theta_l)\mathbf{x}^H\mathbf{A}(\theta_l)\mathbf{x}\right] = 0,
\end{equation}
which gives the optimal $\alpha$ as
\begin{equation}\label{eq:optimal alpha}
\alpha^\star =\frac{\mathbf{x}^H\sum_{l=1}^Ld(\theta_l)\mathbf{A}(\theta_l)\mathbf{x}}{\sum_{l=1}^Ld^2(\theta_l)}.
\end{equation}
Substituting (\ref{eq:optimal alpha}) into $f(\alpha,\mathbf{x})$, the objective (\ref{eq:original objective}) becomes a univariate function\vspace{-0.2 cm}
\begin{equation}
f(\mathbf{x}) = \sum_{l=1}^L\left|\mathbf{x}^H\mathbf{A}_l\mathbf{x}\right|^2,
\end{equation}
where we define
\begin{equation}\label{eq:matrix Al}
\mathbf{A}_l \triangleq \frac{d(\theta_l)\sum_{l=1}^Ld(\theta_l)\mathbf{A}(\theta_l)}{\sqrt{L}\sum_{l=1}^Ld^2(\theta_l)}-\frac{\mathbf{A}(\theta_l)}{\sqrt{L}}, ~\forall l.
\end{equation}
In the meantime, to facilitate the following algorithm development, some basic linear algebra laws are utilized to reformulate the optimization problem in an equivalent concise form:
\begin{subequations}\label{eq:reformulated problem}\begin{align}\label{eq:quartic function}
&\underset{\mathbf{x}}{\min}~~\sum_{l=1}^L\left|\mathbf{x}^H\mathbf{A}_l\mathbf{x}\right|^2 \\
\label{eq:convex constraint}
&~\text{s.t.}~~~\mathfrak{R}\big\{\widetilde{\mathbf{h}}_i^H\mathbf{x}\big\}\geq \gamma_i, i = 1, 2, \ldots, 2K_\text{u},\\
\label{eq:nonconvex constraint}
&~~~~~~~\left|x_m\right| = \sqrt{P_\text{tot}/M}, ~\forall m,
\end{align}
\end{subequations}
where
\begin{subequations}
\begin{align}
\widetilde{\mathbf{h}}_{2k}^H &\triangleq \mathbf{h}_k^H e^{-j\angle s_k}\left(\sin\Phi+ e^{-j\frac{\pi}{2}}\cos\Phi\right), ~\forall k, \\
\widetilde{\mathbf{h}}_{2k-1}^H &\triangleq \mathbf{h}_k^H e^{-j\angle s_k}\left(\sin\Phi- e^{-j\frac{\pi}{2}}\cos\Phi\right), ~\forall k, \\
\gamma_{2k} &\triangleq \beta_k, ~\forall k,~~~~\gamma_{2k-1} \triangleq \beta_k, ~\forall k.
\end{align}
\end{subequations}

It can be seen that the optimization problem (\ref{eq:reformulated problem}) is a non-convex problem due to the quartic objective function (\ref{eq:quartic function}) and the constant-modulus constraint (\ref{eq:nonconvex constraint}), which greatly hinders finding a straightforward solution.
In order to solve these difficulties, in Section \ref{sec:Algorithm 1}, we first utilize the PDD, MM, and BCD methods to convert the original problem into two solvable sub-problems, and then develop efficient algorithms to iteratively solve them.

\section{Proposed PDD-MM-BCD Algorithm}\label{sec:Algorithm 1}
\vspace{0.2 cm}

In this section, we propose a PDD-MM-BCD algorithm to solve the non-convex problem (\ref{eq:reformulated problem}).
In order to tackle the constant-modulus constraint (\ref{eq:nonconvex constraint}), an auxiliary variable $\mathbf{v}$ is first introduced.
Then, the PDD method is applied to handle the coupling constraints and variables, and the MM method is employed to tackle the complicated quartic objective function.
Finally, the BCD method is utilized to iteratively solve each sub-problem.
The details of the algorithm development are described as follows.

\newcounter{TempEqCnt}
\setcounter{TempEqCnt}{\value{equation}}
\setcounter{equation}{17}
\begin{figure*}[!t]
\begin{subequations}\label{eq:linear by MM}
\begin{align}
&\sum_{l=1}^L\left|\mathbf{x}^H\mathbf{A}_l\mathbf{x}\right|^2 \overset{\text{(a)}}{=} \text{vec}^H(\mathbf{xx}^H)\mathbf{B}\text{vec}(\mathbf{xx}^H) \\
&~~~~\overset{\text{(b)}}{\leq}  \lambda_\mathbf{B}\text{vec}^H(\mathbf{xx}^H)\text{vec}(\mathbf{xx}^H) + 2\mathfrak{R}\left\{
\text{vec}^H(\mathbf{xx}^H)(\mathbf{B}-\lambda_\text{B}\mathbf{I}_{M^2})\text{vec}(\mathbf{x}_\text{t}\mathbf{x}_\text{t}^H)\right\}
+ \text{vec}^H(\mathbf{x}_\text{t}\mathbf{x}_\text{t}^H)(\lambda_\text{B}\mathbf{I}_{M^2}-\mathbf{B})\text{vec}(\mathbf{x}_\text{t}\mathbf{x}_\text{t}^H) \\
&~~~~\overset{\text{(c)}}{\leq} \lambda_\mathbf{B}P_\text{tot}^2+\mathbf{x}^H\mathbf{C}\mathbf{x} + \text{vec}^H(\mathbf{x}_\text{t}\mathbf{x}_\text{t}^H)(\lambda_\text{B}\mathbf{I}_{M^2}-\mathbf{B})\text{vec}(\mathbf{x}_\text{t}\mathbf{x}_\text{t}^H)  \\
&~~~~\overset{\text{(d)}}{\leq} \lambda_\mathbf{B}P_\text{tot}^2+ \lambda_\mathbf{C}\mathbf{x}^H\mathbf{x} + 2\mathfrak{R}\left\{\mathbf{x}^H(\mathbf{C}-\lambda_\mathbf{C}\mathbf{I}_M)\mathbf{x}_\text{t}\right\} +
\mathbf{x}_\text{t}^H(\lambda_\mathbf{C}\mathbf{I}_M-\mathbf{C})\mathbf{x}_\text{t} + \text{vec}^H(\mathbf{x}_\text{t}\mathbf{x}_\text{t}^H)(\lambda_\text{B}\mathbf{I}_{M^2}-\mathbf{B})\text{vec}(\mathbf{x}_\text{t}\mathbf{x}_\text{t}^H)   \\
&~~~~\overset{\text{(e)}}{\leq}\mathfrak{R}\left\{\mathbf{x}^H\mathbf{d}\right\} + \varepsilon,
\end{align}
\end{subequations}
\rule[-0pt]{18.5 cm}{0.05em}\vspace{-0.2 cm}
\end{figure*}
\setcounter{equation}{\value{TempEqCnt}}

\subsection{PDD-MM Transformation}
%\vspace{0.1 cm}

In order to decouple the convex constraint (\ref{eq:convex constraint}) and non-convex (\ref{eq:nonconvex constraint}) constraint in variable $\mathbf{x}$, the auxiliary variable $\mathbf{v}\triangleq[v_1, v_2, \ldots, v_M]^T$ is introduced to transform the optimization problem (\ref{eq:reformulated problem}) to\vspace{-0.2 cm}
\begin{subequations}\label{eq:problem xy}
\begin{align}
&\underset{\mathbf{x},\mathbf{v}}{\min}~~\sum_{l=1}^L\left|\mathbf{x}^H\mathbf{A}_l\mathbf{x}\right|^2 \\
&~\text{s.t.}~~~\mathfrak{R}\big\{\widetilde{\mathbf{h}}_i^H\mathbf{x}\big\}\geq \gamma_i, ~\forall i,\\
&~~~~~~~\left|x_m\right| \leq \sqrt{P_\text{tot}/M}, ~\forall m, \\
\label{eq:equality constraint}
&~~~~~~~\mathbf{x} = \mathbf{v}, \\
\label{eq:constant y}
&~~~~~~~\left|v_m\right| = \sqrt{P_\text{tot}/M}, ~\forall m,
\end{align}
\end{subequations}
which is still a non-convex problem due to the coupling variables $\mathbf{x}$ and $\mathbf{v}$ in constraint (\ref{eq:equality constraint}) and the constant-modulus constraint (\ref{eq:constant y}) of variable $\mathbf{v}$.
We then adopt the PDD method to provide a framework for iteratively solving for each variable.
The PDD method is a double-loop algorithm \cite{Shi 17}, in which the inner loop utilizes the BCD approach to iteratively solve the augmented Lagrangian problem, while the outer loop updates the dual variables and/or the penalty parameters.
The details are described next.

By penalizing the equality constraint (\ref{eq:equality constraint}), the augmented Lagrangian problem of (\ref{eq:problem xy}) is written as\setcounter{equation}{18}
\begin{subequations}\label{eq:problem for xy}
\begin{align}
\label{eq:nonconvex objective}
&\underset{\mathbf{x},\mathbf{v}}{\min}~~\sum_{l=1}^L\left|\mathbf{x}^H\mathbf{A}_l\mathbf{x}\right|^2\hspace{-0.1 cm} +\hspace{-0.1 cm} \frac{1}{2\rho}
\left\|\mathbf{x}-\mathbf{v}\right\|^2\hspace{-0.1 cm} + \hspace{-0.1 cm}\mathfrak{R}\{\bm{\mu}^H(\mathbf{x}-\mathbf{v})\}\\
&~\text{s.t.}~~~\mathfrak{R}\big\{\widetilde{\mathbf{h}}_i^H\mathbf{x}\big\}\geq \gamma_i, ~\forall i,\\
\label{eq:xn}
&~~~~~~~\left|x_m\right| \leq \sqrt{P_\text{tot}/M}, ~\forall m, \\
&~~~~~~~\left|v_m\right| = \sqrt{P_\text{tot}/M}, ~\forall m,
\end{align}
\end{subequations}
where $\rho > 0$ is the penalty parameter and $\bm{\mu} \in \mathbb{C}^M$ is the dual variable.
In the inner loop, the BCD algorithm is utilized to iteratively solve for $\mathbf{x}$ and $\mathbf{v}$ with fixed $\rho$ and $\bm{\mu}$.
However, the complicated non-convex objective function (\ref{eq:nonconvex objective}) leads to difficulties in solving each sub-problem.
In order to efficiently address this issue, we use the MM method \cite{Sun TSP 17} and seek for a more tractable surrogate function that locally approximates and upper-bounds the objective (\ref{eq:nonconvex objective}) in each iteration.
The procedure for deriving the surrogate function is described below.

Using a second-order Taylor expansion as in Lemma 12 of \cite{Sun TSP 17}, the surrogate function of a quadratic function at point $\mathbf{x}_\text{t}$ can be constructed as
\begin{equation}\begin{aligned}\label{eq:MM}
\mathbf{x}^H\mathbf{A}_l\mathbf{x} \leq \lambda_{\mathbf{A}_l}\mathbf{x}^H\mathbf{x} &+ 2\mathfrak{R}\left\{\mathbf{x}^H(\mathbf{A}_l-\lambda_{\mathbf{A}_l}\mathbf{I}_M)\mathbf{x}_\text{t}\right\} \\
&+ \mathbf{x}_\text{t}^H(\lambda_{\mathbf{A}_l}\mathbf{I}_M-\mathbf{A}_l)\mathbf{x}_\text{t},
\end{aligned}\end{equation}
where $\lambda_{\mathbf{A}_l}$ is the maximum eigenvalue of the Hermitian matrix $\mathbf{A}_l$.
Thanks to the amplitude constraint (\ref{eq:xn}), the first quadratic term on the right-hand side of (\ref{eq:MM}) is upper-bounded by\vspace{-0.2 cm}
\begin{equation}\label{eq:xx}
\mathbf{x}^H\mathbf{x}\leq P_\text{tot}.
\end{equation}
Substituting (\ref{eq:xx}) into (\ref{eq:MM}), the quadratic function $\mathbf{x}^H\mathbf{A}_l\mathbf{x}$ is upper-bounded by a linear function as
\begin{equation}\begin{aligned}\label{eq:linear MM}
\mathbf{x}^H\mathbf{A}_l\mathbf{x} \leq \lambda_{\mathbf{A}_l}P_\text{tot} &+ 2\mathfrak{R}\left\{\mathbf{x}^H(\mathbf{A}_l-\lambda_{\mathbf{A}_l}\mathbf{I}_M)\mathbf{x}_\text{t}\right\}\\ &+ \mathbf{x}_\text{t}^H(\lambda_{\mathbf{A}_l}\mathbf{I}_M-\mathbf{A}_l)\mathbf{x}_\text{t},
\end{aligned}\end{equation}
which will greatly simplify the optimization and reduce the computational complexity.

Inspired by (\ref{eq:linear MM}), we attempt to transform the first quartic term in objective function (\ref{eq:nonconvex objective}) into a simple linear function by applying the second-order Taylor expansion (\ref{eq:MM}) twice.
The details of the derivations are presented by (\ref{eq:linear by MM}) at the top of the this page, where we define
\begin{subequations}
\begin{align}
\label{eq:matrix B}
\mathbf{B} &\triangleq \sum_{l=1}^L\text{vec}(\mathbf{A}_l)\text{vec}^H(\mathbf{A}_l), \\
\label{eq:matrix C}
\mathbf{C} &\triangleq \text{reshape}\left\{2(\mathbf{B}-\lambda_\mathbf{B}\mathbf{I}_{M^2})\text{vec}(\mathbf{x}_\text{t}\mathbf{x}_\text{t}^H)\right\}_{M\times M}, \\ \mathbf{d} &\triangleq  2(\mathbf{C}-\lambda_\mathbf{C}\mathbf{I}_M)\mathbf{x}_\text{t}, \\
\varepsilon &\triangleq  \lambda_\mathbf{C}P_\text{tot}+
\mathbf{x}_\text{t}^H(\lambda_\mathbf{C}\mathbf{I}_M-\mathbf{C})\mathbf{x}_\text{t} + \lambda_\mathbf{B}P_\text{tot}^2  \\
&\hspace{1.55 cm}+ \text{vec}^H(\mathbf{x}_\text{t}\mathbf{x}_\text{t}^H)(\lambda_\text{B}\mathbf{I}_{M^2}-\mathbf{B})\text{vec}(\mathbf{x}_\text{t}\mathbf{x}_\text{t}^H),
\end{align}
\end{subequations}
and $\lambda_\mathbf{B}$ and $\lambda_\mathbf{C}$ are the maximum eigenvalues of matrices $\mathbf{B}$ and $\mathbf{C}$, respectively.
The $\text{reshape}\{\cdot\}_{M\times M}$ operation represents reshaping the vector to an $M\times M$ dimensional matrix, the $\text{vec}(\cdot)$ operation indicates vectorizing the matrix to a column vector, and $\text{vec}^H(\cdot)$ denotes the conjugate-transpose operation after vectorization.
Substituting (\ref{eq:matrix B}) into (\ref{eq:matrix C}) and utilizing some basic algebra, the matrix $\mathbf{C}$ can be rewritten as $\mathbf{C} = 2\sum_{l=1}^L\mathbf{x}_\text{t}^H\mathbf{A}_l^H\mathbf{x}_\text{t}\mathbf{A}_l
-2\lambda_\mathbf{B}\mathbf{x}_\text{t}\mathbf{x}_\text{t}^H$.
Since matrix $\mathbf{A}_l$ is Hermitian according to its definition in (\ref{eq:matrix Al}), it is easy to confirm that matrices $\mathbf{B}$ and $\mathbf{C}$ are also Hermitian, which enables steps (b) and (d) to utilize (\ref{eq:MM}) to find the upper-bound surrogate function in (\ref{eq:linear by MM}b) and (\ref{eq:linear by MM}d) shown at the top of this page.
With the amplitude constraint (\ref{eq:xn}), step (c) is obtained by \setcounter{equation}{23}
\begin{equation}
\text{vec}^H(\mathbf{x}\mathbf{x}^H)\text{vec}(\mathbf{x}\mathbf{x}^H)\leq P_\text{tot}^2.
\end{equation}
Step (d) is derived using (\ref{eq:xx}).
Similarly, the second quadratic term in (\ref{eq:nonconvex objective}) can be upper-bounded by
\begin{equation}\label{eq:x-y}
\begin{aligned}
\left\|\mathbf{x}-\mathbf{v}\right\|^2 &= \mathbf{x}^H\mathbf{x}+\mathbf{v}^H\mathbf{v}-2\mathfrak{R}\left\{\mathbf{x}^H\mathbf{v}\right\} \\
& \leq P_\text{tot} + P_\text{tot} - 2\mathfrak{R}\left\{\mathbf{x}^H\mathbf{v}\right\}.
\end{aligned}
\end{equation}

\setcounter{TempEqCnt}{\value{equation}}
\setcounter{equation}{31}
\begin{figure*}[!b]
\rule[-0pt]{18.5 cm}{0.05em}
\begin{subequations}\label{eq:dual problem}
\begin{align}
\label{eq:dual problem obj}
&\underset{\bm{\mu},\bm{\nu}}{\min}~~\frac{1}{4}\left(\widehat{\mathbf{H}}^T\bm{\mu}\hspace{-0.1 cm}-\widehat{\mathbf{d}}\right)^T
\left(\sum_{m=1}^M\nu_m\mathbf{E}_m\right)^{-1}\left(\widehat{\mathbf{H}}^T\bm{\mu}-\widehat{\mathbf{d}}\right) -\bm{\mu}^T\bm{\gamma} + \frac{P_\text{tot}}{M}\sum_{m=1}^M\nu_m\\
&~\text{s.t.}~~\bm{\mu} \succeq \mathbf{0},\\
&~~~~~~\bm{\nu} \succeq \mathbf{0},
\end{align}
\end{subequations}
\end{figure*}
\setcounter{equation}{\value{TempEqCnt}}

Therefore, substituting the inequalities in (\ref{eq:linear by MM}) and (\ref{eq:x-y}) into (\ref{eq:nonconvex objective}), the upper-bound surrogate function is given by
\begin{equation}\begin{aligned}
&\sum_{l=1}^L\left|\mathbf{x}^H\mathbf{A}_l\mathbf{x}\right|^2 + \frac{1}{2\rho}
\left\|\mathbf{x}-\mathbf{v}\right\|^2 + \mathfrak{R}\left\{\bm{\mu}^H(\mathbf{x}-\mathbf{v})\right\} \\
& \leq \mathfrak{R}\left\{\mathbf{x}^H\mathbf{d}\right\} + \varepsilon + \frac{P_\text{tot}}{\rho} - \frac{1}{\rho}\mathfrak{R}\left\{\mathbf{x}^H\mathbf{v}\right\} +  \mathfrak{R}\left\{
\bm{\mu}^H(\mathbf{x}-\mathbf{v})\right\}.
\end{aligned}\end{equation}
Then, ignoring the constant term $\varepsilon+P_\text{tot}/\rho$, the optimization problem for variables $\mathbf{x}$ and $\mathbf{v}$ in each iteration can be formulated as
\begin{subequations}\label{eq:problem xy after MM}
\begin{align}
\label{eq:surrogate objective}
&\underset{\mathbf{x},\mathbf{v}}{\min}~~\mathfrak{R}\Big\{\mathbf{x}^H\mathbf{d}-\frac{1}{\rho}\mathbf{x}^H\mathbf{v}+
\bm{\mu}^H(\mathbf{x}-\mathbf{v})\Big\}\\
&~\text{s.t.}~~\mathfrak{R}\big\{\widetilde{\mathbf{h}}_i^H\mathbf{x}\big\}\geq \gamma_i, ~\forall i,\\
&~~~~~~\left|x_m\right| \leq \sqrt{P_\text{tot}/M}, ~\forall m, \\
&~~~~~~\left|v_m\right| = \sqrt{P_\text{tot}/M}, ~\forall m.
\end{align}
\end{subequations}
A two-block BCD algorithm can be utilized to efficiently solve (\ref{eq:problem xy after MM}) by iteratively updating $\mathbf{x}$ and $\mathbf{v}$, as presented in the next subsection.

\subsection{BCD Algorithm}

\textit{Update $\mathbf{v}$}: Ignoring the constant term $\mathfrak{R}\left\{\mathbf{x}^H\left(\mathbf{d}+\bm{\mu}\right)\right\}$ with fixed $\mathbf{x}$, the sub-problem for updating $\mathbf{v}$ is re-arranged as
\begin{subequations}\label{eq:solve for y}
\begin{align}
&\underset{\mathbf{v}}{\max}~~\mathfrak{R}\left\{\left(\mathbf{x}^H+
\rho\bm{\mu}^H\right)\mathbf{v}\right\}\\
&~\text{s.t.}~~\left|v_m\right| = \sqrt{P_\text{tot}/M}, ~\forall m,
\end{align}
\end{subequations}
whose optimal solution can be easily obtained via a phase alignment operation, i.e.,
\begin{equation}\label{eq:optimal y} \mathbf{v}^\star = \sqrt{P_\text{tot}/M}e^{j\angle(\mathbf{x}/\rho+\bm{\mu})}.
\end{equation}

\textit{Update $\mathbf{x}$}: With fixed $\mathbf{v}$, the sub-problem for updating $\mathbf{x}$ is given by\vspace{-0.2 cm}
\begin{subequations}\label{eq:solve for x}\begin{align}
&\underset{\mathbf{x}}{\min}~~\mathfrak{R}\big\{\mathbf{x}^H\widetilde{\mathbf{d}}\big\}\\
&~\text{s.t.}~~\mathfrak{R}\big\{\widetilde{\mathbf{h}}_i^H\mathbf{x}\big\}\geq \gamma_i, ~\forall i,\\
\label{eq:problem x constrait c}
&~~~~~~~\left|x_m\right| \leq \sqrt{P_\text{tot}/M}, ~\forall m,
\end{align}
\end{subequations}
where $\widetilde{\mathbf{d}}\triangleq\mathbf{d}-\mathbf{v}/\rho+\bm{\mu}$ for brevity.
Problem (\ref{eq:solve for x}) is convex and can be solved by various standard methods such as the interior-point method \cite{cvx}.
However, since the double-loop iteration will lead to high computational complexity, a more efficient algorithm is developed in the remainder of this subsection, which solves the Lagrangian dual function of~(\ref{eq:solve for x}) with the aid of the Hook-Jeeves Pattern Search algorithm.

For the algorithm development, we first convert problem (\ref{eq:solve for x}) to an equivalent real-valued form by defining
\begin{equation}\label{eq:definitiona for x}
\begin{aligned}
\overline{\mathbf{x}} &\triangleq \left[\mathfrak{R}\{\mathbf{x}^T\}, \mathfrak{I}\{\mathbf{x}^T\}\right]^T,
\;\;\;\;
\overline{\mathbf{d}} \triangleq \big[\mathfrak{R}\{\widetilde{\mathbf{d}}^T\},  \mathfrak{I}\{\widetilde{\mathbf{d}}^T\}\big]^T,
\\
\overline{\mathbf{h}}_i &\triangleq \big[\mathfrak{R}\{\widetilde{\mathbf{h}}_i^T\}, \mathfrak{I}\{\widetilde{\mathbf{h}}_i^T\}\big]^T,\;\;
\overline{\mathbf{e}}_m \triangleq \big[\mathbf{e}_m^T, \mathbf{0}^T\big]^T, \\
\bm{\Delta}_1 &\triangleq \begin{bmatrix} \mathbf{I}_M & \mathbf{0} \\ \mathbf{0} & -\mathbf{I}_M \end{bmatrix},
\;\;\;\;\;\;\;\;\;\;
\bm{\Delta}_2 \triangleq \begin{bmatrix} \mathbf{0} & \mathbf{I}_M \\ \mathbf{I}_M & \mathbf{0} \end{bmatrix} , \\
\end{aligned}
\end{equation}
where $\overline{\mathbf{e}}_m \in \mathbb{R}^{2M}$ and the auxiliary vector $\mathbf{e}_m\in \mathbb{R}^M$ is a zero-vector except the $m$-th entry is 1.
Then, problem (\ref{eq:solve for x}) is transformed into a real function as\setcounter{equation}{32}
\begin{subequations}\label{eq:real problem for x}
\begin{align}
&\underset{\overline{\mathbf{x}}}{\min}~~\overline{\mathbf{d}}^T\bm{\Delta}_1\overline{\mathbf{x}}\\
\label{eq:linear constraint}
&~\text{s.t.}~~\overline{\mathbf{h}}_i^T\bm{\Delta}_1\overline{\mathbf{x}}\geq \gamma_i, ~\forall i,\\
\label{eq:quadratic constraint}
&~~~~~~ \left|\overline{\mathbf{e}}_m^T\bm{\Delta}_1\overline{\mathbf{x}}\right|^2 + \left|\overline{\mathbf{e}}_m^T\bm{\Delta}_2\overline{\mathbf{x}}\right|^2 \leq P_\text{tot}/M, ~\forall m.
\end{align}
\end{subequations}
To efficiently solve problem (\ref{eq:real problem for x}), we propose to employ its Lagrangian dual function:
\begin{equation}\begin{aligned}\label{eq:Lagrangian dual function}
\mathcal{L}\left(\overline{\mathbf{x}},\bm{\mu},\bm{\nu}\right) = \widehat{\mathbf{d}}^T\overline{\mathbf{x}} &+ \bm{\mu}^T(\bm{\gamma}-\widehat{\mathbf{H}}\overline{\mathbf{x}}) \\
&+ \sum_{m=1}^M\nu_m(\overline{\mathbf{x}}^T\mathbf{E}_m\overline{\mathbf{x}}- \frac{P_\text{tot}}{M}),
\end{aligned}\end{equation}
where
\begin{equation}
\begin{aligned}
\widehat{\mathbf{d}}^T &\triangleq \overline{\mathbf{d}}^T\bm{\Delta}_1, \\
\bm{\gamma} &\triangleq [\gamma_1, \gamma_2, \ldots, \gamma_{2K}]^T, \\
\widehat{\mathbf{H}}&\triangleq \left[\bm{\Delta}_1^T\overline{\mathbf{h}}_1, \;\bm{\Delta}_1^T\overline{\mathbf{h}}_2,\; \ldots, \bm{\Delta}_1^T\overline{\mathbf{h}}_{2K}\right]^T, \\
\mathbf{E}_m &\triangleq \bm{\Delta}_1^T\overline{\mathbf{e}}_m\overline{\mathbf{e}}_m^T\bm{\Delta}_1 + \bm{\Delta}_2^T\overline{\mathbf{e}}_m\overline{\mathbf{e}}_m^T\bm{\Delta}_2,
\end{aligned}
\end{equation}
and $\bm{\mu}\in \mathbb{R}^{2K_\text{u}} \succeq \mathbf{0}$ and $\bm{\nu} \triangleq [\nu_1,\nu_2,\ldots,\nu_M]^T \succeq \mathbf{0}$ are the Lagrangian dual variables.

Setting $\frac{\partial\mathcal{L}}{\partial\overline{\mathbf{x}}}=\mathbf{0}$, the optimal solution $\overline{\mathbf{x}}^\star$ to problem (\ref{eq:real problem for x}) can be calculated as
\begin{equation}\label{eq:optimal xbar}
\overline{\mathbf{x}}^\star = \frac{1}{2}\left(\sum_{m=1}^M\nu_m\mathbf{E}_m\right)^{-1}\left(\widehat{\mathbf{H}}^T\bm{\mu}-\widehat{\mathbf{d}}\right).
\end{equation}
Substituting (\ref{eq:optimal xbar}) into (\ref{eq:Lagrangian dual function}), the Lagrangian dual function is formulated as (\ref{eq:dual problem}), which is presented at the bottom of this page.
Since the Lagrangian dual function~(\ref{eq:dual problem}) has very simple constraints, it can be efficiently solved using standard iterative search algorithms. Given the high complexity required to calculate the first and second order partial derivatives of (\ref{eq:dual problem obj}), we adopt the  derivative-free Hooke-Jeeves Pattern Search algorithm \cite{Liu TIFS 20}, which is a popular local search algorithm whose convergence has been proven in \cite{Torczon SIAMJO 1997}.
The details are omitted due to space limitations.

After obtaining the locally optimal solution $\bm{\mu}^\star$ and $\bm{\nu}^\star$ of the Lagrangian dual problem (\ref{eq:dual problem}), the solution $\overline{\mathbf{x}}^\star$ to the real function (\ref{eq:real problem for x}) can be obtained by (\ref{eq:optimal xbar}).
Then, the solution to the original problem (\ref{eq:solve for x}) is constructed as
\begin{equation}\label{eq:construct x}
\mathbf{x}^\star = \overline{\mathbf{x}}^\star(1:M) + j\overline{\mathbf{x}}^\star(M+1:2M).
\end{equation}

\begin{algorithm}[!t]\begin{small}
\caption{Proposed PDD-MM-BCD Algorithm}
\label{alg:1}
    \begin{algorithmic}[1]
    \REQUIRE $\mathbf{h}_k$, $\beta_k$, $s_k$, $~\forall k$, $d(\theta_l)$, $\mathbf{A}(\theta_l)$, $~\forall \theta_l$, $L$, $\Phi$, $P_\text{tot}$, $\mathbf{E}_m, ~\forall m$, $\bm{\Delta}_1$, $\bm{\Delta}_2$, $0<c<1$, $\delta_\text{{th}}$.
    \ENSURE $\mathbf{x}^\star$, $\alpha^\star$.
        \STATE {Initialize $\mathbf{x}$, $\mathbf{v}$, $\rho$, $\bm{\mu}$.}
        \WHILE {$\left\|\mathbf{x}-\mathbf{v}\right\|_\infty \geq \delta_\text{{th}}$ }
            \STATE {Calculate the objective value $f$ of (\ref{eq:nonconvex objective}).}
            \STATE {Set $\delta := 1$.}
            \WHILE {$\delta \geq \delta_\text{{th}}$}
                \STATE {$f_\text{pre} := f$.}
                \STATE {Update $\mathbf{v}^\star$ by (\ref{eq:optimal y}).}
                \STATE {Calculate $\bm{\mu}^\star$ and $\bm{\nu}^\star$ by Hooke-Jeeves Pattern Search \cite{Liu TIFS 20}.}
                \STATE {Calculate $\overline{\mathbf{x}}^\star$ by (\ref{eq:optimal xbar}).}
                \STATE {Construct $\mathbf{x}^\star$ by (\ref{eq:construct x}).}
                \STATE {Calculate the objective value $f$ of (\ref{eq:nonconvex objective}).}
                \STATE {$\delta := \big|\frac{f-f_\text{pre}}{f}\big|$.}
            \ENDWHILE
            \STATE {Update $\bm{\mu} := \bm{\mu} + (\mathbf{x}-\mathbf{v})/\rho$.}
            \STATE {Update $\rho := c\rho$.}
        \ENDWHILE
        \STATE {Calculate $\alpha^\star$ by (\ref{eq:optimal alpha}).}
        \STATE {Return $\mathbf{x}^\star$ and $\alpha^\star$.}
    \end{algorithmic}
 \end{small}
\end{algorithm}

\subsection{Summary and Analysis}

\textit{Summary}: Based on the above derivations, the proposed PDD-MM-BCD algorithm is straightforward and summarized in Algorithm 1, where $\delta_\text{{th}}$ is the threshold to judge the convergence, and $0<c<1$ is a variable for updating/decreasing the penalty parameter $\rho$.
In the inner loop, we iteratively solve problems (\ref{eq:solve for y}) and (\ref{eq:solve for x}) for updating $\mathbf{x}$ and $\mathbf{v}$ until the objective value converges.
In the outer loop, the penalty parameter $\rho$ and the dual variable $\bm{\mu}$ are updated in steps 14 and 15 until the equality constraint (\ref{eq:equality constraint}) is approximately met.

\textit{Convergence Analysis}: In the inner loop for solving the augmented Lagrangian problem (\ref{eq:problem for xy}), the objective value is non-increasing since each iteration of the MM method generates non-increasing sequences \cite{Sun TSP 17}. In each iteration, the optimal $\mathbf{v}^\star$ is obtained in closed-form and the locally optimal $\mathbf{x}^\star$ is provided by the Hooke-Jeeves Pattern Search algorithm.
Moreover, since the objective value is lower-bounded by $2\sqrt{P_\text{tot}/M}\left\|\bm{\mu}\right\|$, we can conclude that the solution to problem (\ref{eq:problem for xy}) in the inner loop converges.
Therefore, the convergence of the PDD-based algorithm shown in Algorithm 1 can be guaranteed as analyzed in \cite{Shi 17}.

\textit{Computational Complexity Analysis}: In the inner loop, the complexity for updating $\mathbf{v}^\star$ is of order $\mathcal{O}\left(M\right)$, solving problem (\ref{eq:dual problem}) by the Hooke-Jeeves Pattern Search algorithm has complexity of order $\mathcal{O}\left(M^3(2K_\text{u}+M)\right)$, and constructing $\mathbf{x}^\star$ is of order $\mathcal{O}\left(M\right)$.
The complexity to update the penalty and dual variables is of order $\mathcal{O}\left(M\right)$.
Thus, the total complexity to obtain $\mathbf{x}^\star$ is of order $\mathcal{O}\left(N_\text{tot}M^3(2K_\text{u}+M)\right)$, where $N_\text{tot}$ is the total number of iterations.
The total computational complexity to calculate all possible $\mathbf{x}[n]$ is of order $\mathcal{O}\left(\Omega^{K_\text{u}}N_\text{tot}M^3(2K_\text{u}+M)\right)$.
As stated in \cite{Liu TSP 20}, the complexity of the conventional block-level precoding scheme is of order $\mathcal{O}\left(N_\text{tot}K_\text{u}^{6.5}M^{6.5}\right)$.
We can see that for binary-PSK (BPSK) and QPSK modulations and small-scale systems, our proposed symbol-level precoding scheme is even more efficient than the block-level precoding scheme \cite{Liu TSP 20}.
Furthermore, parallel computation can be used to simultaneously calculate the solution for each transmitted signal, which makes the proposed symbol-level precoding scheme more appealing.

\section{Efficient ALM-RBFGS Algorithm}\label{sec:Algorithm 2}
\vspace{0.2 cm}

Although the proposed algorithm in the previous section is efficient for small-scale systems, its computational complexity will become unaffordable as the number of users increases due to the considerable number of iterations required by the MM-based algorithm.
In order to further reduce the computational complexity for applying the proposed symbol-level precoding scheme to large-scale systems, in this section, we propose a more efficient ALM-RBFGS algorithm to solve problem~(\ref{eq:reformulated problem}).
To benefit from existing efficient algorithms designed for unconstrained problems, we first transform this problem into the Riemannian space endowed with the constant-modulus constraint (\ref{eq:nonconvex constraint}), then penalize the inequality constraints (\ref{eq:convex constraint}) in the objective using ALM.
The efficient RBFGS algorithm is further developed to solve this augmented Lagrangian problem.
The details are given below.

\subsection{Riemannian-ALM Transformation}

\setcounter{TempEqCnt}{\value{equation}}
\setcounter{equation}{40}
\begin{figure*}[!t]
\begin{equation}\label{eq:AL function}
\mathcal{L}(\mathbf{x},\rho,\bm{\mu}) = \sum_{l=1}^L\left|{\mathbf{x}}^H\mathbf{A}_l{\mathbf{x}}\right|^2 + \frac{\rho}{2}\sum_{i=1}^{2K_\text{u}}\max\Big\{0,\mu_i/\rho+{\gamma}_i-
\mathfrak{R}\big\{\widetilde{\mathbf{h}}_i^H\mathbf{x}\big\}\Big\}^2.
\end{equation}
\begin{equation}\label{eq:gradient of AL function}
 \nabla g(\mathbf{x}) = 4\sum_{l=1}^L\mathbf{x}^H\mathbf{A}_l\mathbf{x}\mathbf{A}_l\mathbf{x} + \rho\sum_{i=1}^{2K_\text{u}}\left\{
             \begin{array}{lr}
             \mathbf{0}, &\mu_i/\rho+\gamma_i- \mathfrak{R}\big\{\widetilde{\mathbf{h}}_i^H\mathbf{x}\big\} < 0, \\
             \big(\widetilde{\mathbf{h}}_i^H\mathbf{x}-\mu_i/\rho-\gamma_i\big)\widetilde{\mathbf{h}}_i,
              &\mu_i/\rho+\gamma_i- \mathfrak{R}\big\{\widetilde{\mathbf{h}}_i^H\mathbf{x}\big\} \geq 0.
             \end{array}
\right.
\end{equation}
\rule[-0pt]{18.5 cm}{0.05em}
\end{figure*}
\setcounter{equation}{\value{TempEqCnt}}

There are two main categories of algorithms for handling the non-smooth and non-convex constant-modulus constraint (\ref{eq:nonconvex constraint}) on the Euclidean space.
One is for non-convex relaxation-based algorithms, e.g., semi-definite relaxation (SDR) and MM.
However, the SDR-based algorithm cannot be applied to problem (\ref{eq:reformulated problem}) due to the coupled quartic objective function (\ref{eq:quartic function}) and linear constraints (\ref{eq:convex constraint}).
In addition, the MM-based algorithm as described in the previous section and other iterative non-convex relaxation-based algorithms usually require a considerable number of iterations to approximate the original objective function, which may make the computational complexity unaffordable.
The other category is for algorithms based on alternating minimization, which separate each element of the variable vector from the objective function and iteratively solve for it.
However, the complicated quartic objective function greatly hinders the required decomposition.
Therefore, considering these difficulties in handling constraint (\ref{eq:nonconvex constraint}) in Euclidean space, in this subsection we use the geometric structure of the problem to solve~(\ref{eq:reformulated problem}) on the Riemannian space.

According to the definitions in \cite{Huang SIAM 2015}, the constant-modulus constraint (\ref{eq:nonconvex constraint}) forms an $M$-dimensional complex circle manifold:
\begin{equation}
\mathcal{M}_\text{cc} = \big\{\mathbf{x} \in \mathbb{C}^M: x_m^*x_m = P_\text{tot}/M, ~\forall m\big\},
\end{equation}
which is a smooth Riemannian manifold equipped with an inner product defined on the tangent space:
\begin{equation}
T_{\mathbf{x}}\mathcal{M}_\text{cc} = \big\{\mathbf{z} \in \mathbb{C}^M: \mathfrak{R}\left\{\mathbf{z}\odot \mathbf{x}^*\right\} = \mathbf{0}_M\big\}.
\end{equation}
Based on this definition, problem (\ref{eq:reformulated problem}) is re-formulated on the Riemannian space as
\begin{subequations}\label{eq:rereformulated problem}
\begin{align}
&\underset{\mathbf{x}\in \mathcal{M}_\text{cc}}{\min}~~\sum_{l=1}^L\left|\mathbf{x}^H\mathbf{A}_l\mathbf{x}\right|^2 \\
\label{eq:inequality}
&~~\text{s.t.}~~~~\mathfrak{R}\big\{\widetilde{\mathbf{h}}_i^H\mathbf{x}\big\}\geq \gamma_i, i = 1, 2, \ldots, 2K_\text{u}.
\end{align}
\end{subequations}
Then, in order to convert~(\ref{eq:rereformulated problem}) into an unconstrained problem, ALM is used to iteratively minimize the augmented Lagrangian problem and update the penalty parameter and dual variables \cite{Andreani SIAMJO 2007}.
The augmented Lagrangian function $\mathcal{L}(\mathbf{x},\rho,\bm{\mu})$ of problem (\ref{eq:rereformulated problem}) is given by (\ref{eq:AL function}) presented at the top of next page, where $\rho > 0$ is the penalty parameter and $\bm{\mu} \triangleq [\mu_1, \mu_2, \ldots, \mu_{2K_\text{u}}]^T \succeq \mathbf{0}$ is the Lagrangian dual variable.
The update of $\rho$ and $\bm{\mu}$ will be discussed in Sec. IV-C.
In the $p$-th iteration, the augmented Lagrangian problem with fixed penalty parameter $\rho_p$ and dual variable $\bm{\mu}_p$ is formulated as\setcounter{equation}{42}
\begin{equation}\label{eq:AL problem}
\underset{\mathbf{x}\in \mathcal{M}_\text{cc}}{\min}~~~g(\mathbf{x}) \triangleq \mathcal{L}(\mathbf{x},\rho_p,\bm{\mu}_p),
\end{equation}
which is an unconstrained optimization problem on the Riemannian space $\mathcal{M}_\text{cc}$, and can be efficiently solved using the RBFGS algorithm described in the next subsection.

\vspace{-0.4 cm}

\subsection{RBFGS Algorithm}

The Riemannian manifold resembles a Euclidean space at each point, and the gradients of the cost functions, distances, angles, etc., can be measured thanks to its geometry and the Riemannian metric.
Thus, well-developed algorithms for unconstrained problems in Euclidean space can be readily generalized to the Riemannian manifold.
Considering the convergence speed, numerical stability, and computational complexity, we adopt the BFGS algorithm which belongs to the class of quasi-Newton methods, and refer to the BFGS algorithm generalized to the Riemannian manifold as the RBFGS algorithm \cite{Huang SIAM 2015}.

To facilitate the algorithm development, we first calculate the Euclidean gradient of $g(\mathbf{x})$ as in (\ref{eq:gradient of AL function}) shown at the top of this page.
Then, the Riemannian gradient $\text{grad}\;g(\mathbf{x})$ can be obtained by projecting the Euclidean gradient onto its corresponding tangent space as
\begin{equation}\label{eq:grad gx}
\begin{aligned}
\text{grad}\;g(\mathbf{x}) &= \text{Proj}_{\mathbf{x}}\nabla g(\mathbf{x}) \\
& = \nabla g(\mathbf{x}) - \mathfrak{R}\left\{\nabla g(\mathbf{x})\odot\mathbf{x}^*\right\}\odot\mathbf{x}.
\end{aligned}
\end{equation}
Similar to the typical BFGS algorithm, in each iteration the RBFGS algorithm first utilizes the first-order derivative and the Hessian matrix approximation to determine the search direction, then updates the variable with a certain step size, and finally re-calculates the Hessian matrix approximation.
In particular, in the $q$-th iteration of the RBFGS algorithm, the search direction $\bm{\eta}_q \in T_{\mathbf{x}_q}\mathcal{M}_\text{cc}$ is given by the Newton equation as
\begin{equation}\label{eq:Newton equation}
\bm{\eta}_q = -\mathbf{B}_q^{-1}\text{grad}\;g(\mathbf{x}_q),
\end{equation}
where $\mathbf{B}_q$ is the approximation to the Hessian matrix obtained in the previous iteration.
Then, the step size $\alpha_q\in \mathbb{R}$ is chosen by the Armijo backtracking line search method \cite{Huang SIAM 2015}.
Thus, the update of $\mathbf{x}$ is given by
\begin{equation}\label{eq:update x}
\mathbf{x}_{q+1} = \text{Retr}_{\mathbf{x}_q}\left(\alpha_q\bm{\eta}_q\right),
\end{equation}
where $\text{Retr}_{\mathbf{x}_q}\left(\cdot\right)$ is a retraction on $\mathcal{M}_\text{cc}$ so that the update point remains on the manifold.
According to Example 4.1.1 in \cite{Absil book}, the low-complexity retraction on the complex circle manifold $\mathcal{M}_\text{cc}$ is defined as
\begin{equation}
\text{Retr}_{\mathbf{x}_q}\left(\alpha_q\bm{\eta}_q\right) = \sqrt{P_\text{tot}/M}e^{j\angle\left(\mathbf{x}_q+\alpha_q\bm{\eta}_q\right)}.
\end{equation}
Finally, the update of the Hessian matrix approximation $\mathbf{B}$ is calculated by \cite{Huang SIAM 2015}
\begin{equation}\label{eq:update Hessian approximation}
\mathbf{B}_{q+1}\bm{\xi} = \widetilde{\mathbf{B}}_q\bm{\xi} + \frac{\mathbf{y}_q^H\bm{\xi}}{\mathbf{y}_q^H\mathbf{s}_q}\mathbf{y}_q
 - \frac{\mathbf{s}_q^H\widetilde{\mathbf{B}}_q\bm{\xi}}{\mathbf{s}_q^H\widetilde{\mathbf{B}}_q\mathbf{s}_q}\widetilde{\mathbf{B}}_q\mathbf{s}_q, ~\forall \bm{\xi} \in T_{\mathbf{x}_{q+1}}\mathcal{M}_\text{cc},
\end{equation}
where we define
\begin{subequations}
\begin{align}
\mathbf{y}_q &\triangleq \text{grad}\;g(\mathbf{x}_{q+1}) - \text{Trans}_{\alpha_q\bm{\eta}_q}\left(\text{grad}\;g(\mathbf{x}_q)\right), \\
\mathbf{s}_q &\triangleq \text{Trans}_{\alpha_q\bm{\eta}_q}\left(\alpha_q\bm{\eta}_q\right), \\
\widetilde{\mathbf{B}}_q &\triangleq \text{Trans}_{\alpha_q\bm{\eta}_q}\odot\mathbf{B}_q\odot\left(\text{Trans}_{\alpha_q\bm{\eta}_q}\right)^{-1}.
\end{align}
\end{subequations}
The vector transport operation $\text{Trans}_{\alpha_q\bm{\eta}_q}\left(\cdot\right)$ is introduced since two vectors in different tangent spaces cannot be added directly on the Riemannian space.
For the complex circle manifold, the vector transport is defined as
\begin{equation}
\text{Trans}_{\alpha_q\bm{\eta}_q}\left(\bm{\xi}_q\right) = \bm{\xi}_q -
\mathfrak{R}\left\{\bm{\xi}_q^*\odot\left(\mathbf{x}_q+\alpha_q\bm{\eta}_q\right)\right\}\odot\left(\mathbf{x}_q+\alpha_q\bm{\eta}_q\right).
\end{equation}

With the above derivations, the locally optimal solution $\mathbf{x}^\star$ to the augmented Lagrangian problem (\ref{eq:AL problem}) can be obtained by iteratively updating (\ref{eq:update x}) until convergence.
This RBFGS algorithm is summarized in steps $4-13$ of Algorithm 2.

\begin{algorithm}[!t]\begin{small}
\caption{Efficient ALM-RBFGS Algorithm}
\label{alg:2}
    \begin{algorithmic}[1]
    \REQUIRE $\mathbf{h}_k$, $\beta_k$, $s_k$, $~\forall k$, $d(\theta_l)$, $\mathbf{A}(\theta_l)$, $~\forall \theta_l$, $L$, $\Phi$, $P_\text{tot}$, $\delta_\text{{th}}$, $\theta_\rho > 1$, \\
    $\quad\;\; 0< \theta_\varepsilon < 1$, $\rho_\text{max}$, $\mu_\text{max}$.
    \ENSURE $\mathbf{x}^\star$, $\alpha^\star$.
        \STATE {Initialize $\mathbf{x}_0 \in \mathcal{M}_\text{cc}$, $\mathbf{B}_0 := \mathbf{I}_M$, $\rho_0$, $\mu_i^0$, $p := 0$, $\delta_\text{out} > \delta_\text{th}$.}
        \STATE {Obtain the initial point $\mathbf{x}_0^\text{out}$ and $\mathbf{x}_0^\text{in}$ by solving (\ref{eq:initialization}).}
        \WHILE {$\delta_\text{out} > \delta_\text{th}$}
            \STATE {Initialize $q := 0$, $\text{grad}\;g(\mathbf{x}_0)$, $\delta_\text{in} > \delta_\text{th}$.}
            \WHILE {$\delta_\text{in} > \delta_\text{th}$ }
                \STATE {Obtain $\bm{\eta}_q \in \mathcal{M}_\text{cc}$ by (\ref{eq:Newton equation}).}
                \STATE {Calculate the stepsize $\alpha_q$ by Armijo backtracking line search method \cite{Huang SIAM 2015}.}
                \STATE {Update $\mathbf{x}_{q+1}^\text{in}$ by (\ref{eq:update x}).}
                \STATE {Calculate the Riemannian gradient $\text{grad}\;g(\mathbf{x}_{q+1}^\text{in})$ by (\ref{eq:grad gx}).}
                \STATE {Update the Hessian approximation $\mathbf{B}_{q+1}$ by (\ref{eq:update Hessian approximation}).}
                \STATE {$\delta_\text{in} := \left\|\mathbf{x}_{q+1}^\text{in}-\mathbf{x}_q^\text{in}\right\|$.}
                \STATE {$q := q + 1$.}
            \ENDWHILE
            \STATE {$\mathbf{x}_{p+1}^\text{out} := \mathbf{x}_q^\text{in}$.}
            \STATE {$\mu_i^{p+1} := \min\left\{\max\{0, \mu_i^p+\rho_p(\gamma_i-\mathfrak{R}\{\widetilde{\mathbf{h}}_i^H\mathbf{x}_{p+1}^\text{out}\})\}, \mu_\text{max}\right\}$.}
            \STATE {$\varepsilon_i^{p+1} := \max\left\{\gamma_i-\mathfrak{R}\{\widetilde{\mathbf{h}}_i^H\mathbf{x}_{p+1}^\text{out}\},-\frac{\mu_i^p}{\rho_p}\right\}$.}
            \IF {$p = 0$ or $\max\left\{\left|\varepsilon_i^{p+1}\right|, ~\forall i\right\} \leq \theta_\varepsilon\max\left\{\left|\varepsilon_i^{p}\right|, ~\forall i\right\}$}
            \STATE {$\rho_{p+1} := \rho_p$.}
            \ELSE
            \STATE {$\rho_{p+1} := \min\left\{\theta_\rho\rho_p,\rho_\text{max}\right\}$.}
            \ENDIF
            \STATE{$\delta_\text{out} := \left\|\mathbf{x}_{p+1}^\text{out}-\mathbf{x}_p^\text{out}\right\|$.}
            \STATE {$p := p + 1$.}
        \ENDWHILE
        \STATE {$\mathbf{x}^\star := \mathbf{x}_p^\text{out}$.}
        \STATE {Calculate $\alpha^\star$ by (\ref{eq:optimal alpha}).}
        \STATE {Return $\mathbf{x}^\star$ and $\alpha^\star$.}
    \end{algorithmic}
 \end{small}
\end{algorithm}

\subsection{Parameter Update}

In order to alleviate the effect of ill-conditioning and improve the robustness of this algorithm, the update for dual variables $\mu_i \geq 0,~~\forall i$, and penalty parameter $\rho > 0$ should be carefully designed.
Specifically, in the $p$-th iteration of the ALM-RBFGS algorithm, with the obtained solution $\mathbf{x}^\star$ to the augmented Lagrangian problem (\ref{eq:AL problem}), the dual variable $\mu_i \geq 0,~~\forall i$, is updated by
\be
\mu_i^{p+1} = \min\big\{\max\{0, \mu_i^p+\rho_p(\gamma_i-\mathfrak{R}\{\widetilde{\mathbf{h}}_i^H\mathbf{x}^\star\})\}, \mu_\text{max}\big\},
\ee
where $\mu_\text{max}$ is the maximum limit of $\mu_i$ to provide a good safeguard.
Similarly, $\rho$ has the maximum limit $\rho_\text{max}$.
Defining $\varepsilon_i = \max\big\{\gamma_i-\mathfrak{R}\{\widetilde{\mathbf{h}}_i^H\mathbf{x}^\star\},-\frac{\mu_i}{\rho}\big\}$ as the violation to the $i$-th constraint, the maximum violation to all constraints can be expressed as  $\max_i\{|\varepsilon_i|\}$.
The penalty parameter $\rho$ will be updated as
\be
\rho_{p+1} = \min\{\theta_\rho\rho_p,\rho_\text{max}\}
\ee
with $\theta_\rho > 1$ only when the constraint violations shrink fast enough, i.e.,
\be
\max_i\{|\varepsilon_i^{p+1}|\} \geq \theta_\varepsilon\max_i\{|\varepsilon_i^p|\},
\ee
where $\theta_\varepsilon$ represents the parameter to evaluate the shrinkage speed; otherwise $\rho$ retains the same value as in the previous iteration.
The update of the dual variables $\mu_i$ and the penalty parameter $\rho$ is summarized as shown in steps $15-21$ of Algorithm 2.

\subsection{Summary and Analysis}

\textit{Summary}: With above derivations, the procedure for the proposed ALM-RBFGS algorithm is straightforward and summarized in Algorithm 2.
It is noted that the RBFGS algorithm is a local search method, thus a warm-start is preferable.
Intuitively, the solution to the scenario that does not consider communication services is a great choice.
The optimization problem for initialization is thus expressed as
\begin{equation}\label{eq:initialization}
\underset{\mathbf{x}\in \mathcal{M}_\text{cc}}{\min}~~\sum_{l=1}^L\left|\mathbf{x}^H\mathbf{A}_l\mathbf{x}\right|^2,
\end{equation}
which can be efficiently solved by the RBFGS algorithm as in steps $4-12$ with a random feasible initial point $\mathbf{x}_0\in \mathcal{M}_\text{cc}$.
Given an initial point by solving problem (\ref{eq:initialization}), the solution $\mathbf{x}^\star$ to the augmented Lagrangian problem (\ref{eq:AL problem}), the penalty parameter $\rho$, and the dual variable $\bm{\mu}$ are iteratively updated until convergence is achieved.

\textit{Computational Complexity Analysis}:
The computational complexity of using the RBFGS algorithm to update $\mathbf{x}$ is of order $\mathcal{O}\left\{M^3\right\}$, and the computational complexity of updating $\bm{\mu}$ and $\rho$ is of order $\mathcal{O}\left\{4K_\text{u}M\right\}$.
Therefore, the total computational complexity of the proposed ALM-RBFGS algorithm is of order $\mathcal{O}\left\{\Omega^{K_\text{u}}N_\text{tot}\left(M^3+4K_\text{u}M\right)\right\}$.
It can be seen that the ALM-RBFGS algorithm is theoretically more efficient than the PDD-MM-BCD algorithm.
Moreover, the numerical results in the next section will show that the ALM-RBFGS algorithm requires much fewer iterations and less execution time, which makes it a more practical implementation for large-scale systems.

\section{Extensions for Doppler Processing}
\label{sec:extend}
\vspace{0.2 cm}

As mentioned previously, the approach presented in this paper focuses only on the spatial properties of the radar, and assumes the Doppler component is negligible.
When the targets of interest are rapidly moving, Doppler effects must be taken into account, and the temporal characteristics of the radar waveform come into play.
Given the symbol-level nature of the communication constraints, this requires non-trivial extensions to the approaches proposed herein.
In this section, we briefly discuss two possible approaches that could be taken to tackle this problem.

Standard radar implementations employ temporal pulse sequences whose correlation properties are carefully chosen to trade-off resolution in range and Doppler.
In MIMO radar, these become sequences of vector pulses with further considerations of spatial correlation, which leads to designs that also must take into account the spatial response (e.g., such as the beampattern as considered here).
Since symbol-level precoding is by definition focused on transmission of a single symbol, modifications are necessary when designing symbol sequences.

One approach would be to use the idea of \textit{waveform similarity}, which quantifies the difference between the transmitted waveform and some ideal reference waveform with desirable space-time correlation properties.
In particular, let $\widetilde{\mathbf{x}}_0\triangleq [\widetilde{\mathbf{x}}_0^T[1],\ldots,\widetilde{\mathbf{x}}_0^T[N]]^T\in\mathbb{C}^{NM}$ represent the reference waveform, where $N$ denotes the number of radar pulses.
The waveform optimization problem could then be formulated as
\begin{subequations}\label{eq:problem extension}\begin{align}
&\underset{\mathbf{x}[1],\ldots,\mathbf{x}[N]}{\min}~~\sum_{n=1}^N\big\|\mathbf{x}[n]-\widetilde{\mathbf{x}}_0[n]\big\|^2 \\
&\hspace{0.6 cm}\text{s.t.}\hspace{0.8 cm}\mathfrak{R}\big\{\widetilde{\mathbf{h}}_i^H\mathbf{x}[n]\big\}\geq \gamma_i, ~~\forall i,~n,\\
&\hspace{1.8 cm}\left|x_m[n]\right| = \sqrt{P_\text{tot}/M},~~\forall m,~n.
\end{align}
\end{subequations}
This problem can be equivalently divided into $N$ sub-problems, each of which has a form similar to the one given in problem (15), and thus could be solved using the proposed algorithms with some straightforward modifications.

Instead of using a reference waveform, a second approach would be to again pose the problem as one of precoding the entire space-time waveform as in~\eqref{eq:problem extension}, but with an objective function and constraints that address the dual radar-communications functionality.
As an example, for the communications centric approach considered in this paper, the radar objective could involve matching a desired angle-Doppler spectrum or ambiguity function on the radar side, and the communication constraints could be imposed on the symbols over the entire block.
In this case, the ``symbol''-level precoding would involve a super-symbol that is decoded as a block rather than individually (for a discussion of implementing symbol-level precoding over a block of data, see \cite{JeddaMSN18}).
This would further allow the corresponding constructive regions to account for built-in temporal redundancy due to channel coding as well, since certain codewords would not be valid.

The details of extensions such as those discussed above are beyond the scope of this paper, but they serve to emphasize the generality of the symbol-level precoding idea to DFRC systems.

\vspace{0.3 cm}
\section{Simulation Results}\label{sec:simulation results}

In this section, we provide extensive simulation results to evaluate the performance of the proposed symbol-level precoding designs for DFRC systems.
To demonstrate the advantages of our proposed symbol-level precoding algorithms, the state-of-the-art block-level precoding approaches for DFRC systems in \cite{Liu TWC 18} and \cite{Liu TSP 20} are included for comparison.
The instantaneous transmit beampatterns are first plotted in Sec.~\ref{sec:simulation results}-A to intuitively illustrate the advantages of the proposed schemes in radar sensing.
Then, the radar sensing and multi-user communication performance is quantitatively evaluated in Sec.~\ref{sec:simulation results}-B to demonstrate the superiority of the proposed symbol-level precoding schemes.
Finally, the comparisons of convergence and computational complexity are shown in Sec.~\ref{sec:simulation results}-C to illustrate the efficiency of the proposed algorithms.
The simulations are carried out using Matlab 2020b on a PC with an Intel Core i7-9700 CPU and 32 GB of RAM.

In the simulations, we assume that the noise power and QoS requirements for each communication user are the same, i.e., $\sigma^2 = \sigma_k^2 = 10 \text{dBm},$ $\beta = \beta_k, ~\forall k$.
For a fair comparison with the conventional block-level precoding schemes, in which the SINR threshold is $\Gamma$, we set the QoS requirement of our proposed schemes as $\beta = \sigma\sin\Phi\sqrt{\Gamma}$.
The following simulation will show that this setting makes the communication constraint (\ref{eq:fig2 communication constraint}) stricter than the block-level precoding counterpart.
The noise power in the received radar signal is $\sigma^2_\text{z} = 20 \text{dBm}$, and the total transmit power is set as $P_\text{tot} = 30 \text{dBm}$.
The BS is equipped with $M = 10$ antennas with antenna spacing $\Delta = \lambda/2$. QPSK constellation symbols are assumed for all users.
We assume $K_\text{t} = 3$ targets at the locations $\theta_1 = -40\degree$, $\theta_2 = 0\degree$, and $\theta_3 = 40\degree$, respectively, with the same amplitudes $\beta_\theta = 1$.
The ideal beampattern is thus given by
\begin{equation}
d(\theta) = \left\{
\begin{aligned}
1 & , \;\;\; \theta_i - \frac{\Delta_\theta}{2} \leq \theta \leq \theta_i + \frac{\Delta_\theta}{2}, i = 1, 2, 3, \\
0 & , \;\;\; \text{otherwise},
\end{aligned}
\right.
\end{equation}
where $\Delta_\theta = 10\degree$ is the beam width in the desired directions.
The direction grids $\left\{\theta_l\right\}_{l=1}^L$ are uniformly sampled from $-90\degree$ to $90\degree$ with a resolution of $1\degree$.
The scaling parameters are set as $c = 0.8$, $\theta_\rho = 1.1$, and $\theta_\epsilon = 0.6$.
In addition, the convergence threshold $\delta_\text{th}$ is set as $10^{-5}$ in the simulations.

\subsection{Instantaneous Transmit Beampattern}

\begin{figure}[!t]\centering
\subfigure[Conventional block-level precoding scheme \cite{Liu TSP 20}.]{
\begin{minipage}[b]{0.5\textwidth}
\centering
\includegraphics[width = 3.5 in]{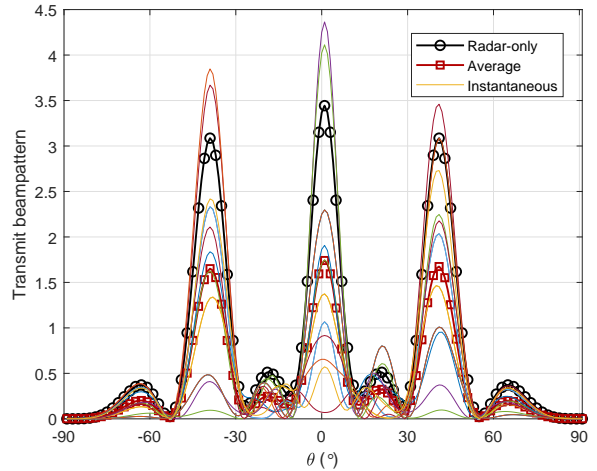}\vspace{0.1 cm}
\end{minipage}
}
\subfigure[Proposed symbol-level precoding scheme (PDD-MM-BCD).]{
\begin{minipage}[b]{0.5\textwidth}
\centering
\includegraphics[width = 3.5 in]{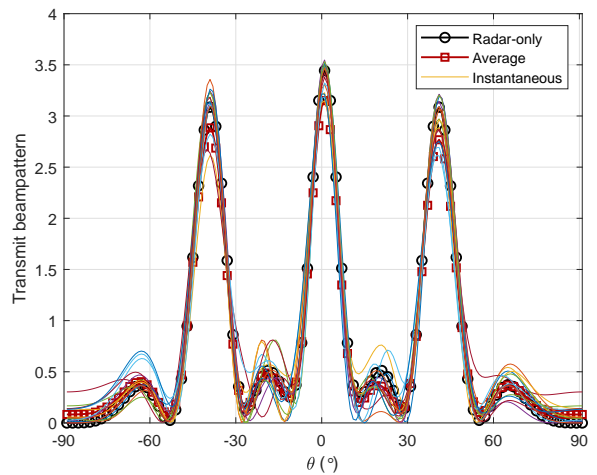}\vspace{0.1 cm}
\end{minipage}
}\vspace{0.1 cm}
\caption{Instantaneous transmit beampatterns (Black lines represent the radar-only benchmark \cite{Stoica TSP 2007}, other  colored lines represent the obtained beampatterns in different time slots except for the red lines with square markers denoting their average).}
\label{fig:beampattern}
\end{figure}

The instantaneous transmit beampatterns are shown in Fig.~\ref{fig:beampattern} for the block-level precoding scheme in \cite{Liu TSP 20} and the proposed symbol-level precoding approach using the PDD-MM-BCD algorithm, where black lines with circle markers represent the radar-only benchmark \cite{Stoica TSP 2007}, and other colored lines denote the obtained transmit beampatterns in different time slots, except for the red lines with square markers which denote their average. The QoS requirement for the $K_\text{u} = 3$ communication users is set as $\Gamma = 6\text{dB}$.
It is clear that the instantaneous transmit beampatterns of the proposed symbol-level precoding scheme in Fig.~\ref{fig:beampattern}(b) always maintain satisfactory similarity with the ideal beampattern in all time slots; i.e., they are all clustered around the radar-only benchmark, while the block-level precoding counterparts have dramatic fluctuations as shown in Fig.~\ref{fig:beampattern}(a).
Furthermore, we can observe that the average transmit beampattern of these snapshots in Fig.~\ref{fig:beampattern}(b) for symbol-level precoding is also better than that in Fig.~\ref{fig:beampattern}(a) for block-level precoding.
Thus, we can conclude that the symbol-level precoding technique provides significantly better consistency in generating well-formed beampatterns for each time slot and can achieve better average transmit beampatterns with limited samples.
In addition to the qualitative results in Fig.~\ref{fig:beampattern}, quantitative evaluations of the target detection and parameter estimation performance will be shown in the next subsection to verify the advantages of the proposed symbol-level precoding scheme.

\subsection{Comparisons of Radar and Communication Performance }

\begin{figure}[!t]
\centering
\includegraphics[width = 3.5 in]{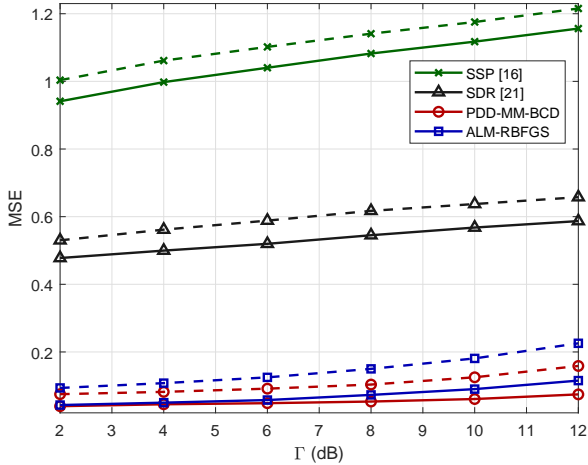}\vspace{0.1 cm}
\caption{Transmit beampattern MSE versus the QoS requirements of multi-user communication $\Gamma$ (Solid lines represent the $K_\text{u} = 3$ scenario, and dashed lines represent the $K_\text{u} = 4$ scenario).}
\label{fig:MSE}
\end{figure}

In this subsection, we first evaluate the radar performance in terms of transmit beampattern MSE in Fig. \ref{fig:MSE}.
The transmit beampattern MSE is defined as the average squared error between the optimal radar-only transmit beampattern \cite{Stoica TSP 2007} and the obtained instantaneous transmit beampattern:
\begin{equation}
\text{MSE} = \mathbb{E}\left\{\frac{1}{L}\sum_{l=1}^L\Big|\mathbf{a}^H(\theta_l)\mathbf{R}^\star\mathbf{a}(\theta_l)-
\mathbf{x}^H[n]\mathbf{A}(\theta_l)\mathbf{x}[n]\Big|^2\right\},
\end{equation}
where $\mathbf{R}^\star$ is the optimal covariance matrix for the radar-only scenario.
In Fig. \ref{fig:MSE}, the MSE versus the communication QoS requirements $\Gamma$ is plotted for $K_\text{u} = 3$ (solid lines) and $K_\text{u} = 4$ (dashed lines) scenarios, where ``SSP \cite{Liu TWC 18}'' and ``SDR \cite{Liu TSP 20}'' respectively denote the block-level precoding schemes solved by the sum-square penalty based algorithm \cite{Liu TWC 18} and the SDR-based algorithm \cite{Liu TSP 20}, ``PDD-MM-BCD'' and ``ALM-RBFGS'' denote the proposed symbol-level precoding designs solved by Algorithm 1 and Algorithm 2, respectively.
Not surprisingly, the MSE for all schemes increases with larger $\Gamma$ and $K_\text{u}$, which shows the performance trade-off between radar sensing and multi-user communications.
In addition, the proposed symbol-level precoding algorithms can dramatically reduce the transmit beampattern MSE compared with the conventional block-level precoding schemes, which results primarily for the following two reasons: \textit{i)} The proposed symbol-level precoding approaches focus on the instantaneous transmit beampatterns by designing the transmitted signal in each time slot, while the block-level precoding methods \cite{Liu TWC 18}, \cite{Liu TSP 20} only consider the average transmit beampattern by optimizing the second-order statistics of the transmitted signals.
\textit{ii)} Our developed non-linear symbol-level precoding designs can exploit more DoFs than the linear block-level precoding approach.
We also observe that the performance of SDR \cite{Liu TSP 20} is better than that of SSP \cite{Liu TWC 18}, since the former scheme transmits both the precoded communication symbols and the radar waveform, which can exploit more DoFs for radar sensing than the latter scheme.
Furthermore, compared with the PDD-MM-BCD algorithm, only a slight performance loss is observed for the ALM-RBFGS algorithm, and the loss grows as $\Gamma$ increases since stronger penalty terms to guarantee higher communication QoS requirements affect the optimization of the radar performance objective.

\begin{figure}[!t]
\centering
\includegraphics[width = 3.5 in]{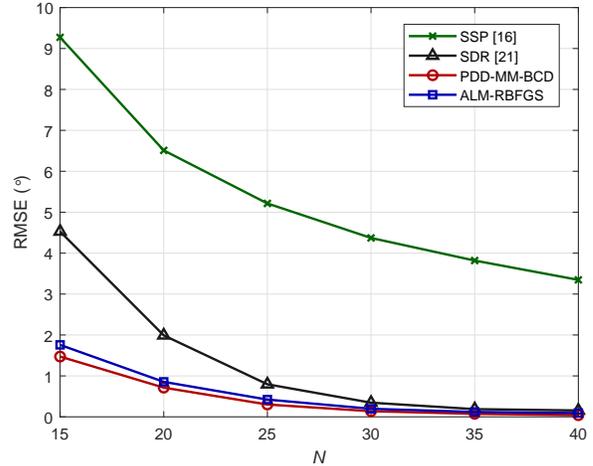}\vspace{0.2 cm}
\caption{Angle estimation RMSE versus the number of collected signals $N$ ($K_\text{u} = 3$ and $\Gamma = 6\text{dB}$).}
\label{fig:RMSE}%\vspace{-0.2 cm}
\end{figure}

\begin{figure}[!t]
\centering
\includegraphics[width = 3.5 in]{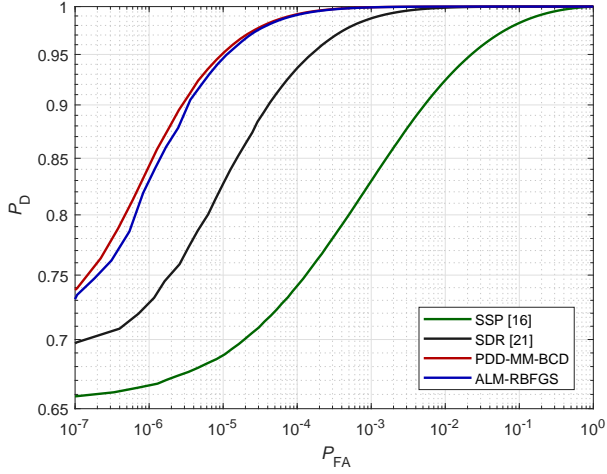}\vspace{0.2 cm}
\caption{Detection probability versus false alarm probability ($K_\text{u} = 3$, $\Gamma = 6\text{dB}$, and $N = 15$).}
\label{fig:ROC}
\end{figure}

Next, in order to illustrate the advantages of the proposed symbol-level precoding designs in guaranteeing preferable target angular estimation performance with limited radar samples, we show the target angular estimation performance versus the number of collected samples in Fig.~\ref{fig:RMSE}.
The popular generalized likelihood ratio test (GLRT) \cite{XU TAES 2008} is used to process the collected signals and estimate the angles $\{\widehat{\theta}_{k_\text{t}}\}_1^{K_\text{t}}$.
Before providing a brief review of the GLRT method, we first define $\mathbf{X} \triangleq [\mathbf{x}[1], \ldots, \mathbf{x}[N]]$, $\mathbf{Y} \triangleq[\mathbf{y}[1], \ldots, \mathbf{y}[N]]$, $\mathbf{Z} \triangleq [\mathbf{z}[1], \ldots, \mathbf{z}[N]]$, where $N$ is the number of collected samples, and $\mathbf{r}^H(\theta) \triangleq \mathbf{a}^H(\theta)\mathbf{X}$.
The typical GLRT method is developed to test if there exists a target at the angular location $\theta$, thus the following hypothesis testing problem is considered
\begin{equation}
\left\{\begin{array}{lr}H_0: \mathbf{Y} = \mathbf{Z},&\\
H_1: \mathbf{Y} = \beta_\theta\mathbf{a}(\theta)\mathbf{r}^H(\theta)+\mathbf{Z}.&
\end{array}
\right.
\end{equation}
The generalized-likelihood ratio (GLR) at the angular direction $\theta$ is defined by
\begin{equation}\label{eq:rho GLRT}
\rho_\text{G}(\theta) = 1-\left[\frac{\max f(\mathbf{Y}|H_0)}{\max f(\mathbf{Y}|H_1)}\right]^{1/N},
\end{equation}
where $f(\mathbf{Y}|H_i)$ is the probability density function (PDF) of $\mathbf{Y}$ under the hypothesis $H_i,~i = 0,1$.
According to (\ref{eq:rho GLRT}), the GLR at each $\theta \in [-90^\circ,90^\circ]$ with a certain resolution can be calculated.
Then, the angular direction of a target can be estimated by
\be\label{eq:estimate theta 1}
\widehat{\theta}_1 = \arg\max_{\theta}~\rho_\text{G}(\theta).
\ee
For the case with multiple targets, the iterative GLRT (iGLRT) \cite{XU TAES 2008} is used to estimate the angular directions of the $K_\text{t}$ targets as follows.
Suppose we have detected and located $\kappa$ targets at the angles $\{\theta_{k_\text{t}}\}_1^\kappa$, and we are testing if there is a $(\kappa+1)$-th target. The corresponding hypothesis testing problem is
\begin{equation}
\left\{\begin{array}{lr}H_\kappa:~~~\mathbf{Y} = \sum_{k_\text{t}=1}^{\kappa}\beta_{k_\text{t}}\mathbf{a}(\theta_{k_\text{t}})\mathbf{r}^H(\theta_{k_\text{t}}) + \mathbf{Z},\\
H_{\kappa+1}: \mathbf{Y} = \beta_\theta\mathbf{a}(\theta)\mathbf{r}(\theta)+\sum_{k_\text{t}=1}^{\kappa}\beta_{k_\text{t}}\mathbf{a}(\theta_{k_\text{t}})
\mathbf{r}^H(\theta_{k_\text{t}}) + \mathbf{Z},
\end{array}
\right.
\end{equation}
and the corresponding conditional GLR (cGLR) \cite{XU TAES 2008} is defined by
\be\label{eq:rho iGLRT}
\rho_\text{G}(\theta|\{\widehat{\theta}_{k_\text{t}}\}_1^\kappa) = 1-\left[\frac{\max f(\mathbf{Y}|H_\kappa)}{\max f(\mathbf{Y}|H_{\kappa+1})}\right]^{1/N}.
\ee
With the cGLR in (\ref{eq:rho iGLRT}), the angle of the ($\kappa+1$)-th target is estimated by
\be\label{eq:conditional estimated angle}
\widehat{\theta}_{\kappa+1} = \arg\max_{\theta}~\rho_\text{G}(\theta|\{\widehat{\theta}_{k_\text{t}}\}_1^\kappa).
\ee
Detailed derivations and expressions for $\rho_\text{G}(\theta)$ in (\ref{eq:rho GLRT}) and $\rho_\text{G}(\theta|\{\widehat{\theta}_{k_\text{t}}\}_1^\kappa)$ in (\ref{eq:rho iGLRT}) can be found in \cite{XU TAES 2008} and are omitted here for brevity. In summary, we first use (\ref{eq:estimate theta 1}) to detect the first target and estimate its angle.
Then, the remaining $K_\text{t}-1$ targets are conditionally detected and estimated by (\ref{eq:conditional estimated angle}) in sequence.

The target angular estimation performance is evaluated in terms of the root-mean-square-error (RMSE) of the estimated target angles, which is defined as
\begin{equation}
\text{RMSE} = \sqrt{\mathbb{E}\left\{\frac{1}{K_\text{t}}\sum_{{k_\text{t}}=1}^{K_\text{t}}
\left(\theta_{k_\text{t}}-\widehat{\theta}_{k_\text{t}}\right)^2\right\}},
\end{equation}
where $\theta_{k_\text{t}}$ is the actual angle and $\widehat{\theta}_{k_\text{t}}$ is the estimated angle of the $k_\text{t}$-th target.
From Fig. \ref{fig:RMSE}, we see that as the number of collected samples increases, the estimation accuracy improves for all schemes.
However, our proposed algorithms always provide better angular estimation performance than \cite{Liu TWC 18} and \cite{Liu TSP 20} since the non-linear symbol-level precoding designs can exploit more DoFs than the linear block-level precoding counterparts, despite the fact that they also impose stricter communication constraints.
It is also worth noting that with very few collected samples, e.g., $N = 15$, the proposed schemes can provide much lower RMSE, which reveals the potentials of the symbol-level precoding designs in fast radar sensing cases.

\begin{figure}[!t]
\centering
\includegraphics[width = 3.5 in]{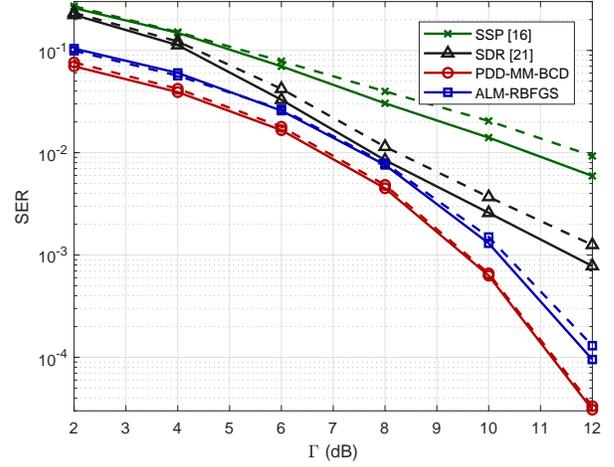}\vspace{0.2 cm}
\caption{Average SER versus the QoS requirements of multi-user communication $\Gamma$ (Solid lines represent the $K_\text{u} = 3$ scenario, and dashed lines represent the $K_\text{u} = 4$ scenario).}
\label{fig:SER}%\vspace{-0.2 cm}
\end{figure}

In order to evaluate the target detection performance, we present the receiver operating characteristic (ROC) by plotting the detection probability $P_\text{D}$ versus the false alarm probability $P_\text{FA}$ in Fig. \ref{fig:ROC}.
In particular, the ($\kappa+1$)-th target is detected by testing
\be
\rho_\text{G}(\theta|\{\widehat{\theta}_{k_\text{t}}\}_1^\kappa) \underset{H_\kappa}{\overset{H_{\kappa+1}}{\gtrless}} \delta,
\ee
where $\delta$ is the threshold which determines $P_\text{FA}$ and $P_\text{D}$.
In addition, it can be observed that with the same false alarm probability, e.g., $10^{-5}$, both the proposed PDD-MM-BCD and ALM-RBFGS algorithms achieve much higher detection probability than the two block-level approaches, which illustrates the superiority of the proposed symbol-level precoding scheme for target detection.

Finally, we evaluate the communication performance in terms of SER in Fig. \ref{fig:SER}, where the same settings as in Fig. \ref{fig:MSE} are assumed.
Since the communication constraints of the symbol-level precoding algorithms are stricter than those of the block-level precoding methods, a lower SER for the symbol-level approaches is observed in Fig.~\ref{fig:SER}.
Moreover, since the constraints of the communication QoS are incorporated into the objective function as a penalty term in \cite{Liu TWC 18}, they cannot always be satisfied in the optimizations, which causes worse communication performance, i.e., higher SER.
In addition, we see that the SER performance of the ALM-RBFGS algorithm is worse than that for the PDD-MM-BCD algorithm since violations of the communication constraints exist in optimizing the augmented Lagrangian problem (\ref{eq:AL problem}).
However, the slight performance loss is acceptable especially considering the significant computational complexity reduction as illustrated in the next subsection. Furthermore, the ALM-RBFGS algorithm still provides better SER performance than the block-level precoding approaches, which makes it a very competitive candidate for DFRC systems.

\begin{figure}[!t]
\centering
\subfigure[Inner loop.]{
\begin{minipage}{0.22\textwidth}
\centering
\includegraphics[width = 1.8 in]{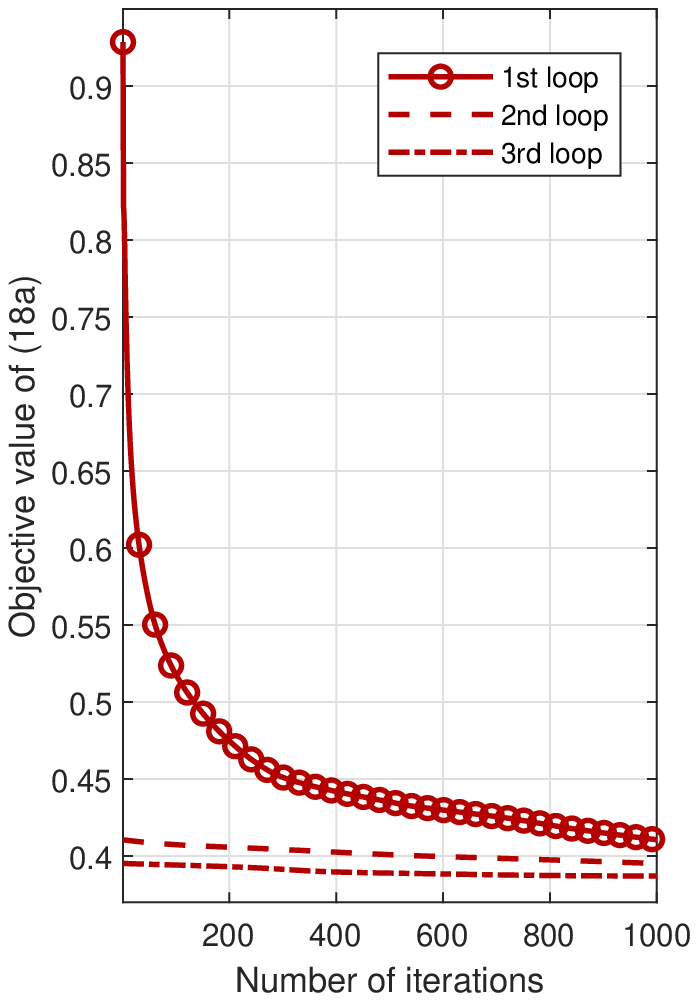}\vspace{0.2 cm}\label{fig:iteration11}
\end{minipage}
}
\subfigure[Outer loop.]{
\begin{minipage}{0.22\textwidth}
\centering
\includegraphics[width = 1.8 in]{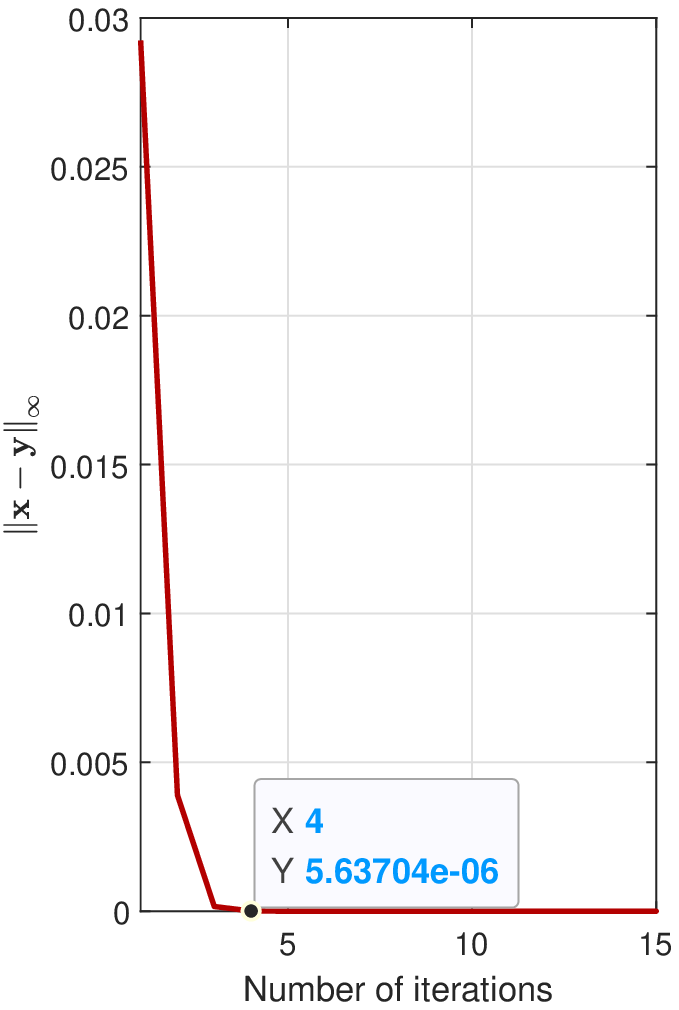}\vspace{0.2 cm}\label{fig:iteration12}
\end{minipage}
}
\vspace{0.2 cm}
\caption{Convergence of PDD-MM-BCD algorithm.}
\label{fig:convergence1}\vspace{-0.2 cm}
\end{figure}

\subsection{Comparisons of Convergence and Complexity}

The convergence performance of the proposed PDD-MM-BCD algorithm is illustrated in Fig. \ref{fig:convergence1}, where the QoS requirement for the $K_\text{u} = 3$ communication users is set as $\Gamma = 6\text{dB}$.
Fig. \ref{fig:iteration11} shows the convergence of the objective value (\ref{eq:nonconvex objective}) in the inner loop, where the curve with circle markers denotes the initial loop.
It can be seen that the inner loop monotonically converges within a limited number of iterations, and the required number of iterations sharply decreases after the initial loop.
The convergence of the outer loop is presented in Fig.~\ref{fig:iteration12}, where we see that the error of the equality constraint~(\ref{eq:equality constraint}) quickly converges within 4 iterations.
Furthermore, with the aid of the more efficient Hooke-Jeeves Pattern Search algorithm rather than the CVX toolbox, the computational complexity of the entire PDD-MM-BCD algorithm is manageable for moderate-scale systems.

\begin{figure}[!t]
\centering
\subfigure[Inner loop.]{
\begin{minipage}{0.22\textwidth}
\centering
\includegraphics[width = 1.8 in]{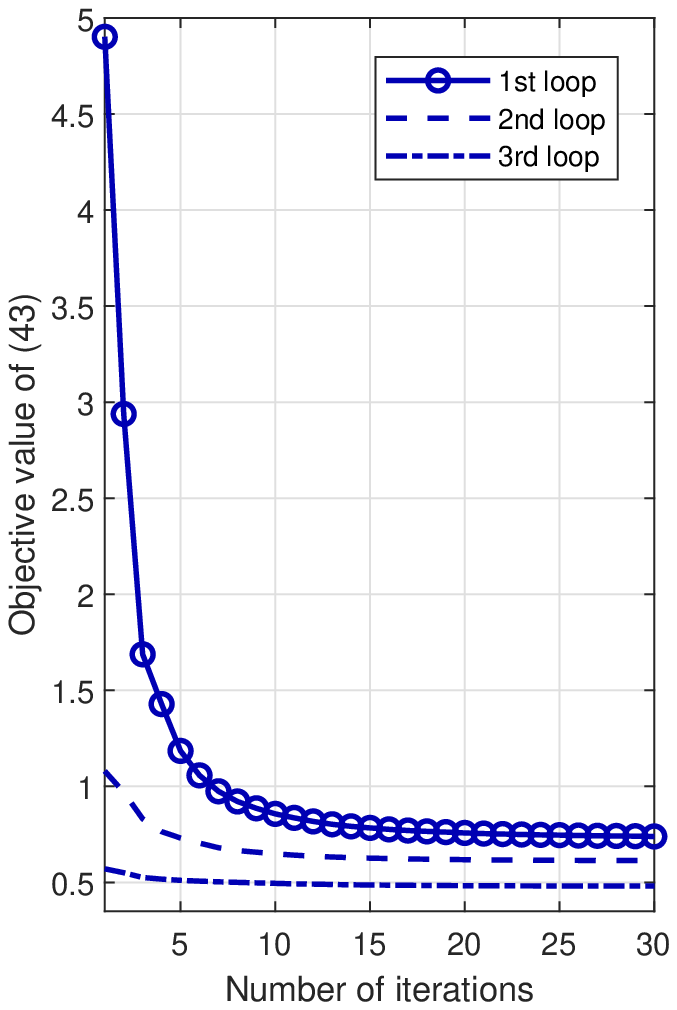}\label{fig:iteration21}\vspace{0.2 cm}
\end{minipage}
}
\subfigure[Outer loop.]{
\begin{minipage}{0.22\textwidth}
\centering
\includegraphics[width = 1.8 in]{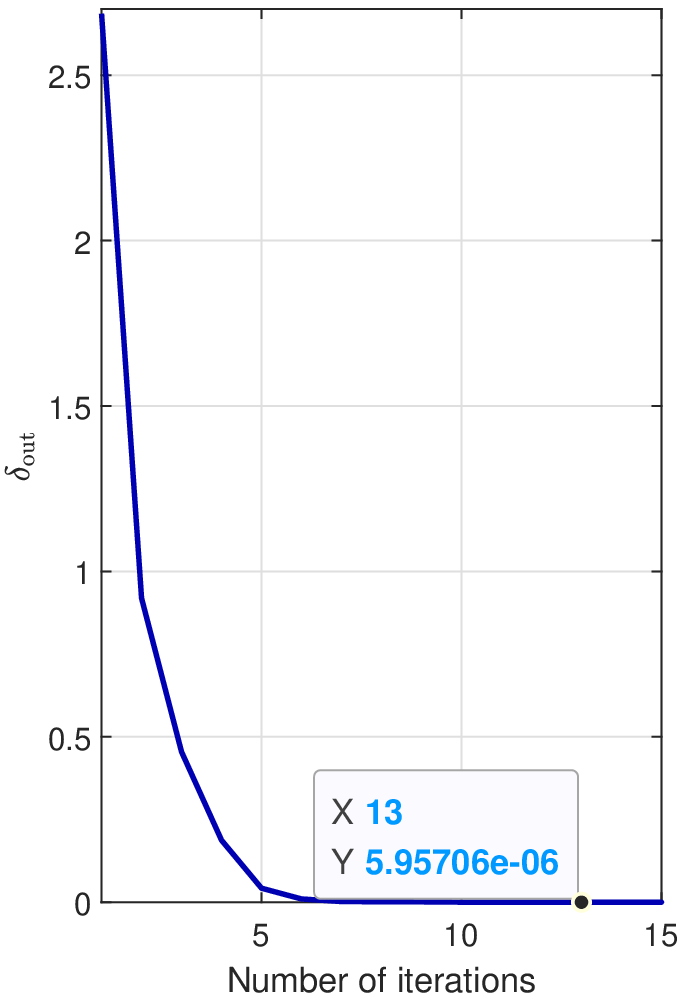}\label{fig:iteration22}\vspace{0.2 cm}
\end{minipage}
}\vspace{0.2 cm}
\caption{Convergence of ALM-RBFGS algorithm.}\label{fig:convergence2}
\vspace{0.2 cm}
\end{figure}

The convergence performance of the proposed ALM-RBFGS algorithm is shown in Fig.~\ref{fig:convergence2}, where the settings are the same as in Fig.~\ref{fig:convergence1}.
In particular, Fig.~\ref{fig:convergence2}(a) plots the convergence of the objective value~(\ref{eq:AL problem}) obtained by the RBFGS algorithm, and Fig.~\ref{fig:convergence2}(b) presents the convergence of ALM in the outer loop, where $\delta_\text{out}$ is the difference of two consecutive iterations as defined in step 22 of Algorithm 2.
The convergence of both the inner and outer loops is very rapid.
Moreover, comparing Fig. \ref{fig:convergence2}(a) with Fig. \ref{fig:convergence1}(a), we can easily conclude that the ALM-RBFGS algorithm requires orders of magnitude fewer iterations in solving the augmented Lagrangian problem.
This is because the ALM-RBFGS algorithm directly solves the quartic objective function rather than iteratively approximating it as in the PDD-MM-BCD algorithm.
Thus, the total computational complexity of the ALM-RBFGS algorithm will be much lower than the PDD-MM-BCD algorithm.

\begin{small}
\begin{table}[!t]
\centering
\caption{\label{tab:time1}Average execution time (seconds) to obtain the precoded transmit vector.}
\vspace{-0.2 cm}
\begin{center}
\begin{tabular}{ c | c | c | c | c | c }
   \hline
   $K_\text{u}$     & 2     & 3     & 4     & 5     & 6   \\
   \hline
   PDD-MM-BCD     &5.57    &7.52   &8.37    &8.47   &9.91   \\
   \hline
   ALM-RBFGS    &0.149    &0.158    &0.189    &0.203    &0.233   \\
   \hline
\end{tabular}
\end{center}
\vspace{-0.2 cm}
\end{table}
\end{small}

Finally, to provide a more intuitive and direct comparison of complexity, the average execution time required to obtain the precoded transmit vector using the PDD-MM-BCD and ALM-RBFGS algorithms is presented in Table I.
We see that the execution time increases as the number of users increases, since the increasing number of constraints results in a larger number of variables and more iterations.
Comparing the two proposed algorithms, it can be seen that the ALM-RBFGS approach is much more efficient and only requires only about $2\%$ of the execution time of the PDD-MM-BCD method, which makes the performance loss shown in the previous subsections acceptable.
Moreover, in practical implementations, parallel computation can be applied to pre-calculate the possible transmitted signals, which will further significantly reduce the execution time.

\section{Conclusions}\label{sec:conclusions}
\vspace{0.2 cm}

In this paper, we introduced the novel symbol-level precoding technique to DFRC systems and investigated the associated waveform designs.
The squared error between the designed and desired beampatterns was minimized while satisfying symbol-level constraints on the communication QoS and constant-modulus power. We proposed two efficient algorithms to solve this non-convex problem with different trade-offs in terms of performance and complexity. Simulation results show that the proposed symbol-level precoding approach provides more accurate angle estimation and better target detection performance with limited collected signals, as well as lower SER for multi-user communications compared with conventional block-level precoding methods. These results reveal the immense potential of symbol-level precoding in DFRC systems. Motivated by this initial work, more complicated symbol-level precoding based DFRC systems deserve further investigation, e.g., in hostile radar sensing environments with the presence of clutter or jamming signals. In addition to the considered beampattern squared error metric, the work can be extended to use other radar waveform design metrics such as mutual information, SINR, the Cram\'{e}r-Rao bound, etc., to explore the performance improvements in radar sensing offered by symbol-level precoding. Extensions to the case with non-negligible Doppler as discussed in Section~V also warrant further attention.

\end{document}